\documentclass[final]{cvpr}

\usepackage{times}
\usepackage{float}
\usepackage{epsfig}
\usepackage{graphicx}
\usepackage{subfigure}
\usepackage{amsmath}
\usepackage{amssymb}

\usepackage[pagebackref=true,breaklinks=true,colorlinks,bookmarks=false]{hyperref}

\def\cvprPaperID{5236} % *** Enter the CVPR Paper ID here
\def\confYear{CVPR 2021}

\begin{document}

%%%%%%%%% TITLE
\title{A Cross Channel Context Model for Latents in Deep Image Compression}

\author{Changyue Ma, Zhao Wang, Ruling Liao, Yan Ye\\
Alibaba Group\\
{\tt\small changyue.mcy, baixiu.wz, ruling.lrl, yan.ye@alibaba-inc.com}
%\and
%Zhao Wang\\
}

\maketitle

%%%%%%%%% ABSTRACT
\begin{abstract}
This paper presents a cross channel context model for latents in deep image compression. Generally, deep image compression is based on an autoencoder framework, which transforms the original image to latents at the encoder and recovers the reconstructed image from the quantized latents at the decoder. The transform is usually combined with an entropy model, which estimates the probability distribution of the quantized latents for arithmetic coding. Currently, joint autoregressive and hierarchical prior entropy models are widely adopted to capture both the global contexts from the hyper latents and the local contexts from the quantized latent elements. For the local contexts, the widely adopted 2D mask convolution can only capture the spatial context. However, we observe that there are strong correlations between different channels in the latents. To utilize the cross channel correlations, we propose to divide the latents into several groups according to channel index and code the groups one by one, where previously coded groups are utilized to provide cross channel context for the current group. The proposed cross channel context model is combined with the joint autoregressive and hierarchical prior entropy model. Experimental results show that, using PSNR as the distortion metric, the combined model achieves BD-rate reductions of 6.30\% and 6.31\% over the baseline entropy model, and 2.50\% and 2.20\% over the latest video coding standard Versatile Video Coding (VVC) for the Kodak and CVPR CLIC2020 professional dataset, respectively. %To the best of our knowledge, our approach is the first work to achieve superior performance over VVC in both low and high resolution test images in terms of PSNR.
In addition, when optimized for the MS-SSIM metric, our approach generates visually more pleasant reconstructed images.
\end{abstract}

%%%%%%%%% BODY TEXT
\section{Introduction}

\begin{figure}[t]
\begin{center}
\includegraphics[width=0.96\linewidth]{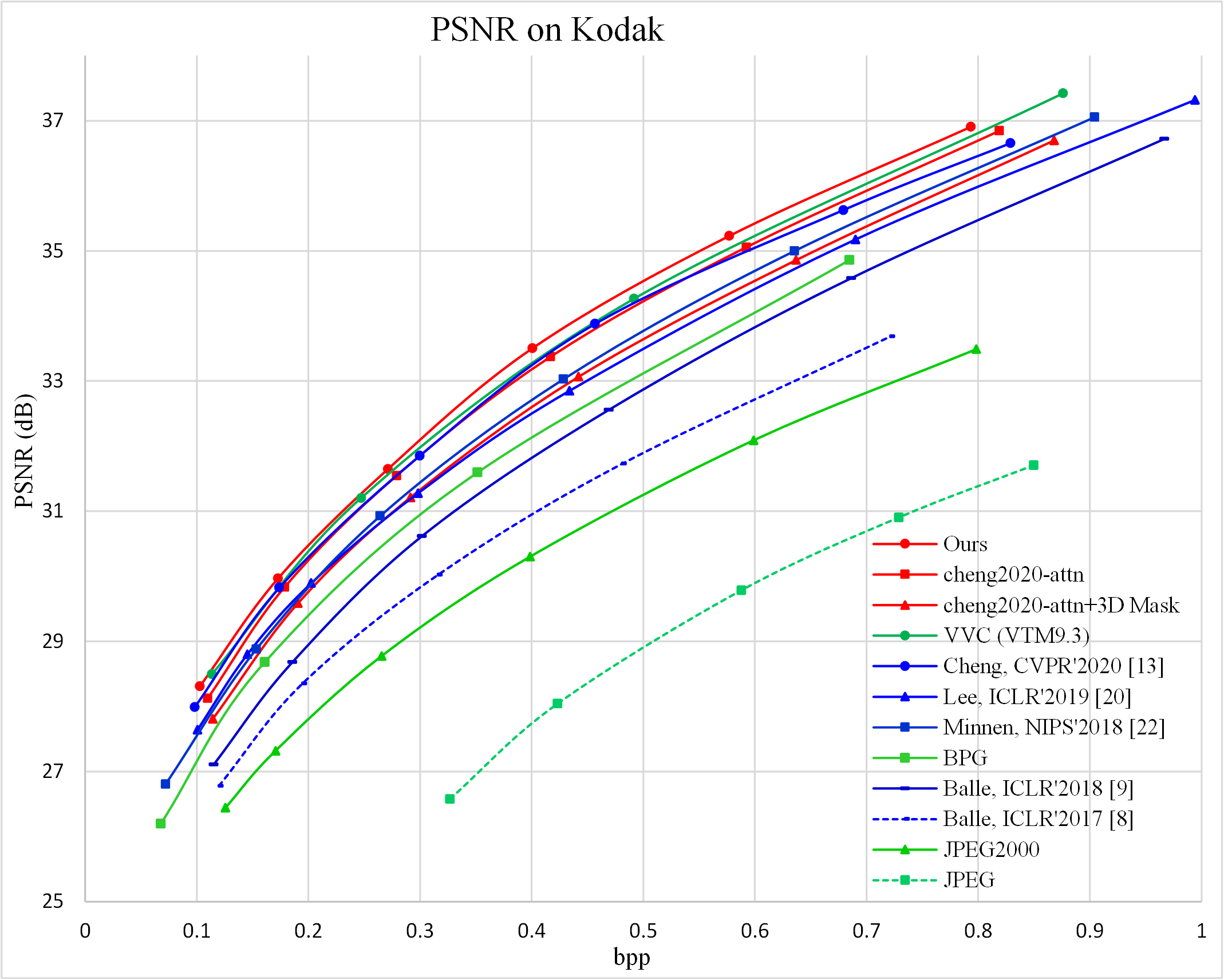}
\end{center}
\caption{Performance comparison on Kodak dataset with PSNR distortion metric.}
\label{fig:KodakPSNR}
\end{figure}

Image compression plays a critical role in image storage and transmission systems.  Over the past decades, a large number of companies and institutions around the world have been working on image compression
and released several image coding standards, such as the widely used JPEG~\cite{wallace1992jpeg} and JPEG2000~\cite{rabbani2002overview} standards. Further, recent video coding standards such as the H.265/HEVC~\cite{sullivan2012overview} and the recently finalized H.266/VVC~\cite{vvc2020} also support efficient image compression with their respective still image profiles. All of these standards use a hybrid coding framework which includes block-based intra prediction, transform, quantization and entropy coding to exploit the spatial, visual and statistical redundancy in images to achieve efficient compression. However, as the modules in the hybrid coding framework are usually designed individually, it becomes more and more difficult to further improve the coding performance based on the basic framework.

Recently, deep image compression has exhibited a fast developing trend with promising results. Compared to traditional image compression methods, deep image compression can optimize all the modules in its compression framework in an end-to-end manner. Currently, among all the deep image compression methods, transform coding together with a context-adaptive entropy model~\cite{balle2016end,balle2018variational,minnen2018joint,cheng2020learned} is the most representative approach to achieve the best coding performance. In ~\cite{balle2016end}, Ballé \emph{et al.} proposed a basic deep image compression framework which makes all steps in the framework differentiable. After that, the initial factorized entropy model is improved by introducing a Gaussian mixture model with the mean and scale parameters estimated by combining the hyperprior and an autoregressive mask convolution~\cite{balle2018variational,minnen2018joint,cheng2020learned}. In addition, the network structure of the transform is modified by introducing the residual connection and the attention module~\cite{cheng2020learned}. Our work is based on these new technologies and further improves the coding performance.

In this paper, we propose a cross channel context model for latents in deep image compression. This is motivated by our observation that there are strong correlations among different channels in the latents currently not utilized by the widely adopted 2D mask convolution method. Although the work~\cite{chen2019neural} attempts to use 3D mask convolution to capture both spatial and cross channel correlation, the coding performance of the 3D mask convolution is not satisfactory. The reason may be because correlations among adjacent channels may not be as strong since there are no constraints on the ranking of different channels when the deep image compression framework is trained. To more effectively utilize the cross channel correlation, we divide the latents into several groups according to channel index and encode them one by one. When coding one group, all previously encoded groups are utilized to provide the cross channel context for the current group, which can build a wider range of correlation across multiple channels. 

In addition, the authors in ~\cite{li2020learning}, ~\cite{lee2019end} propose to combine the global spatial context and local spatial context for better entropy estimation, which is different from our paper to utilize cross channel context and local spatial context. The global spatial context for each channel can only utilize the further up or left encoded latents as context. However, cross channel context can utilize the co-located, down or right encoded latents as context. The authors in ~\cite{minnenchannel} propose the basic idea of the cross channel context model. However, there are some new features in our paper. First, the core motivation of our paper comes from the observation that although there are strong correlations among different channels, the strongest correlations may not always reside between adjacent channels, that is why we need to combine all previously encoded channels to extract cross channel context. In addition, we observe that cross channel context and local spatial context are complementary so they are combined in our paper. As a comparison, the authors in ~\cite{minnenchannel} propose to replace local spatial context with cross channel context to reduce complexity and only cross channel context is utilized in ~\cite{minnenchannel}. Then, the implementation platform and specific network structures of the proposed method are different from ~\cite{minnenchannel}.   

The proposed cross channel context model is integrated into the \emph{cheng2020-attn} model in the deep image compression platform CompressAI~\cite{CompressAI}. Experimental results show that, based on the PSNR distortion metric, the proposed method achieves BD-rate reductions of 6.30\% and 6.31\% over the baseline entropy model, and 2.50\% and 2.20\% over VVC test model VTM9.3 for the Kodak and CVPR CLIC2020 professional datasets, respectively. %To the best of our knowledge, it is the first time that deep image compression method has superior coding performance over VVC in both low and high resolution test images regarding PSNR.
In addition, when the proposed method is optimized with the MS-SSIM distortion metric, visually more  pleasant reconstructed results are obtained.
%-------------------------------------------------------------------------
\section{Related Work}

\textbf{Conventional Image Compression} Existing image compression standards such as JPEG~\cite{wallace1992jpeg}, JPEG 2000~\cite{rabbani2002overview}, or still image profiles of video compression standards, such as H.265/HEVC~\cite{sullivan2012overview}, VVC~\cite{vvc2020}, are based on the hybrid coding framework. In this framework, the input image is first split into several non-overlapping blocks. Then, each block successively goes through intra prediction, discrete cosine/sine transform or wavelet transform, quantization, and Huffman coding or context adaptive binary arithmetic coding (CABAC) to generate the final bitstreams. After decoding the reconstructed images, loop filters can be used to further enhance the reconstruction quality.  Compared with previous standards, the latest coding standard VVC supports more flexible block partition structure, more intra prediction modes, more transform kernels and so on, and achieves superior coding performance. 
%In addition, along with the development of image coding standards, some hybrid approaches which combines traditional image coding framework with deep learning technology have been proposed, typical examples include deep learning based loop filter~\cite{dai2017convolutional} and deep learning based arithmetic coding~\cite{ma2019convolutional}. 

\textbf{Deep Image Compression} In recent years, we have witnessed the fast development of deep image compression technologies. 
%Generally speaking, deep image compression is based on an autoencoder structure. At encoder, the original image is transformed to latents. After quantization, the quantized latents are compressed into bit stream with an entropy coder. At decoder, the reconstructed image is obtained by inverse transforming the quantized latents decoded from the bit stream. During the development of deep image compression, many related works have been proposed to improve the coding performance. 
At the beginning, research focused on solving the basic problems of making the quantization and rate estimation steps differentiable in deep image compression framework~\cite{balle2016end,theis2017lossy,agustsson2017soft}. Later, researchers proposed to use the Gaussian mixture model to replace the initial factorized model, where the mean and scale parameters are estimated by combining the hyperprior and autoregressive information \cite{balle2018variational,minnen2018joint,lee2018context,cheng2020learned}. Besides, advanced network structures of residual connection~\cite{he2016deep}, attention model~\cite{hu2018squeeze}, non-local structure~\cite{wang2018non}, and so on are integrated into the transform network to generate more compact latents~\cite{cheng2020learned,chen2019neural}. In addition to the work above, which are based on convolutional neural network, some work~\cite{toderici2015variable,toderici2017full,johnston2018improved} adopted recurrent neural network to compress the residuals progressively (though they mainly rely on the binary representation at each iteration to achieve scalable coding), and other work~\cite{rippel2017real,santurkar2018generative,agustsson2019generative} used generative adversarial networks to compress the images at extremely low bit rates and pursue the subject quality (though it is difficult to guarantee the objective quality). However, when all related approaches are taken into account, there is still a coding performance gap between deep image compression and the latest coding standard VVC.

\textbf{Context Model} Context model has a long history during the development of deep image compression technologies. At the beginning, based on the basic idea of PixelRNN~\cite{oord2016pixel}, the authors in~\cite{mentzer2018conditional} proposed to estimate the probability of each latent based on previously coded latents with 3D mask convolution. Later, joint autoregressive and hierarchical prior entropy models were proposed to capture both the global and the local context models~\cite{minnen2018joint,lee2018context,cheng2020learned}, where the local context is captured by 2D mask convolution. Considering 2D mask convolution can not capture the cross channel correlation, the authors in~\cite{chen2019neural} proposed to integrate 3D mask convolution into the joint model to capture both the spatial and the cross channel contexts. However, the coding performance is not satisfactory as adjacent channels may not have strong correlations.

\section{Proposed Method}

\subsection{Formulation of Deep Image Compression}
% 0.175, 0.235, 0.235, 0.235
% 0.15, 0.215, 0.215, 0.215
\begin{figure*}[ht]
\begin{center}
\subfigure[Basic model~\cite{balle2016end}]{
\includegraphics[width=0.175\linewidth]{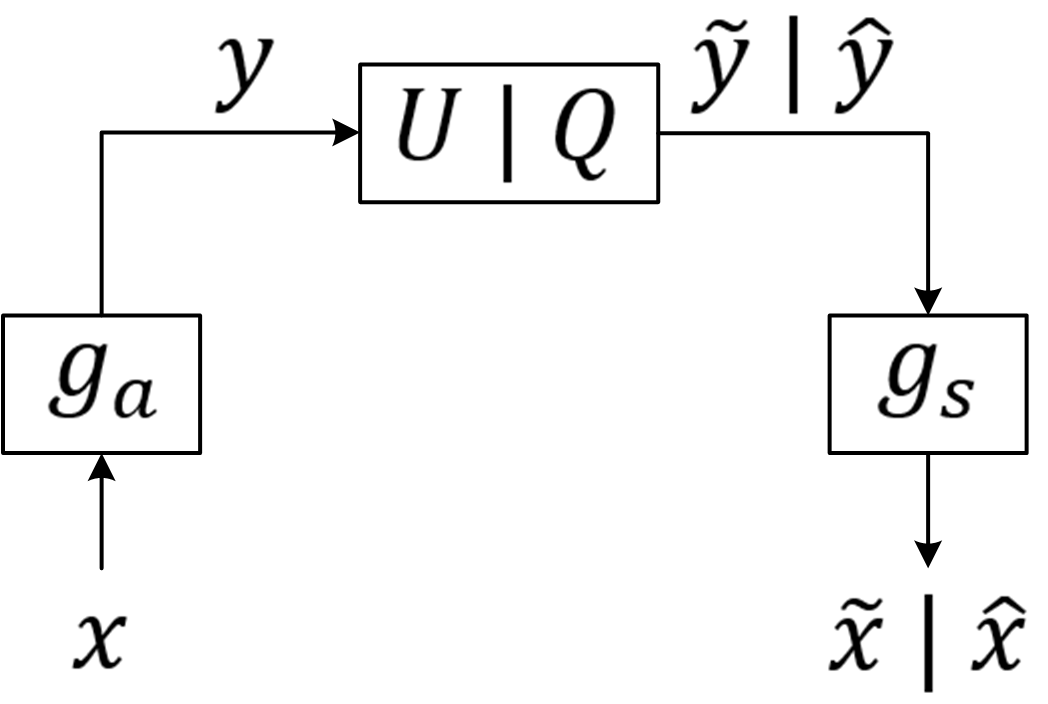}
\label{fig:fw1}
}
\subfigure[Scale hyperprior~\cite{balle2018variational}]{
\includegraphics[width=0.235\linewidth]{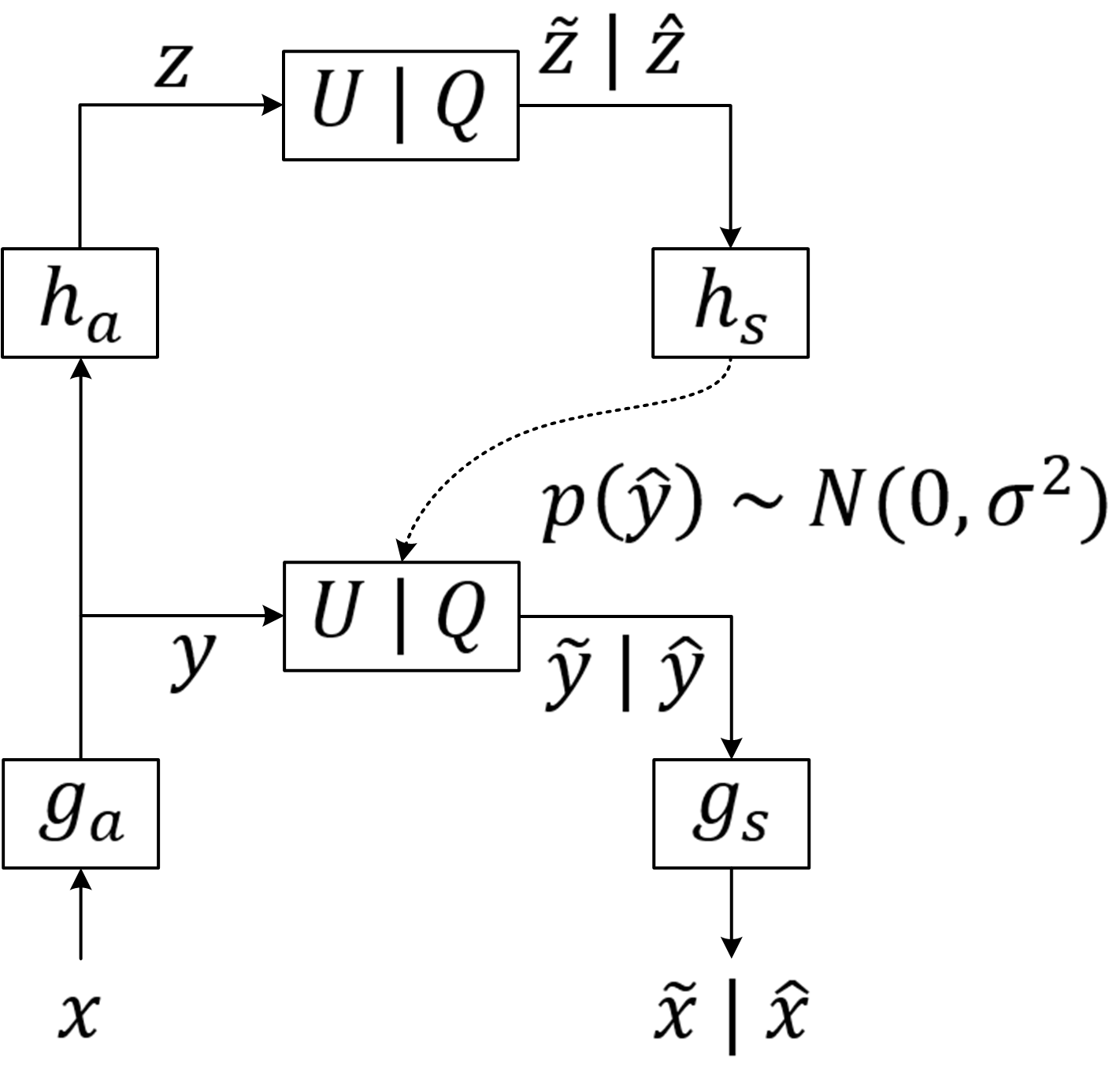}
\label{fig:fw2}
}
\subfigure[Joint model~\cite{minnen2018joint}]{
\includegraphics[width=0.235\linewidth]{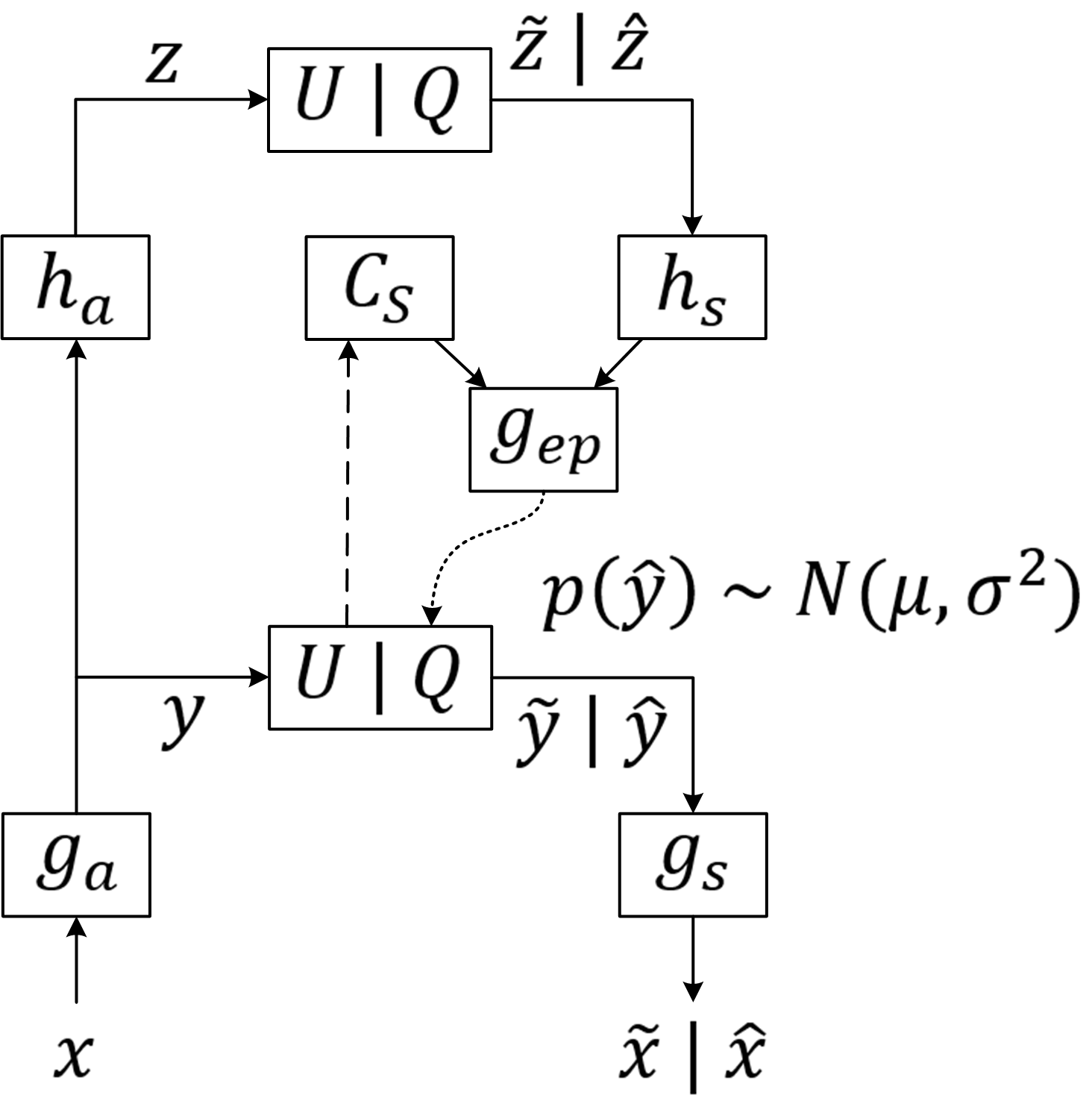}
\label{fig:fw3}
}
\subfigure[The proposed model]{
\includegraphics[width=0.235\linewidth]{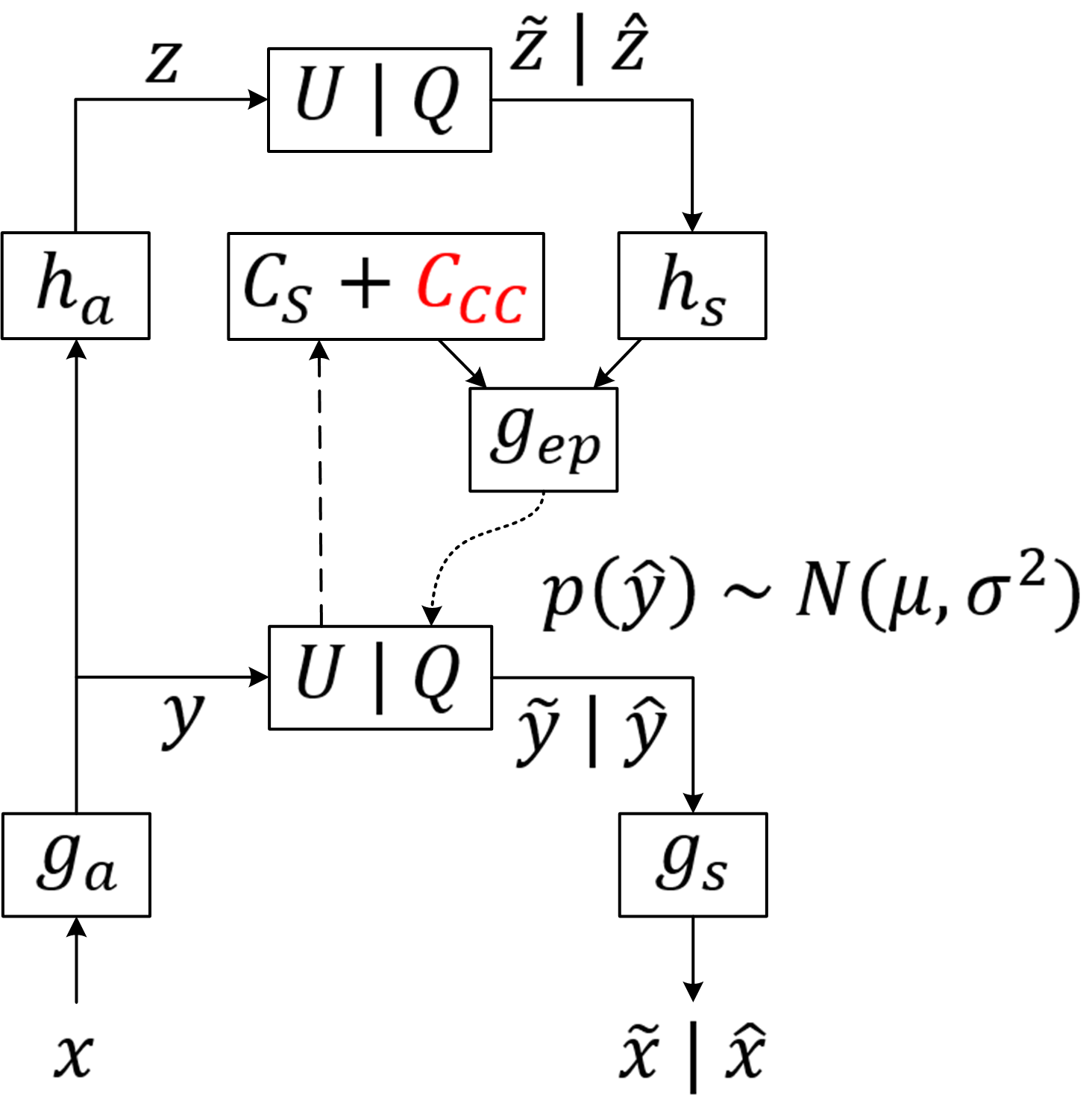}
\label{fig:fw4}
}
\end{center}
   \caption{Operational diagrams of deep image compression models (a)(b)(c) and the proposed cross channel context model (d).}
\label{fig:framework}
\end{figure*}

In~\cite{balle2016end}, the authors proposed the basic deep image compression framework. As shown in Fig. \ref{fig:fw1}, for the original image, it is first mapped to latents via a parameterized analysis transform, then the latents are quantized and compressed into bit stream. After decoding the quantized latents from the bit stream, the reconstructed image is obtained  from the quantized latents via a parameterized synthesis transform. Above process is summarized as follows:
\begin{equation}
\begin{split}
y &= g_a(x;\phi) \\
\hat{y} &= Q(y) \\
\hat{x} &= g_s(\hat{y};\theta)
\end{split}
\end{equation}
where $x$ and $\hat{x}$ represent the original and reconstructed image, $y$ and $\hat{y}$ represent the latents before and after quantization, $g_a$ and $g_s$ represent the analysis and synthesis transform with parameters $\phi$ and $\theta$, and $Q(.)$ represents the quantization process. When training the end-to-end framework, to make the quantization differentiable, the authors in~\cite{balle2016end} propose to add uniform noise ($U$) to simulate the quantization step, and the corresponding quantized latents and reconstructed image are represented with $\tilde{y}$ and $\tilde{x}$. When testing, the rounding operation is used to quantize the latents. For entropy coder, a factorized model is utilized to estimate the probability distribution of the latents.

Following the basic deep image compression framework, the authors in~\cite{balle2018variational} proposed to model the probability distribution of the latents with a zero mean Gaussian model, where the scale parameter of the Gaussian distribution is estimated by a hyperprior. The hyperprior is a side information generated from the latents and needs to be encoded into bitstream. As shown in Fig. \ref{fig:fw2}, the generation process of the hyperprior is summarized as follows:
\begin{equation}
\begin{split}
z &= h_a(y;{\phi}_h) \\
\hat{z} &= Q(z) \\
p_{\hat{y}|\hat{z}} & \sim N(0,{\sigma}^2) \\
 with \  \sigma &= h_s(\hat{z};{\theta}_h) 
\end{split}
\end{equation}
where $z$ and $\hat{z}$ represent the hyper latents before and after quantization, and $h_a$ and $h_s$ represent the hyper analysis and synthesis transform with parameters ${\phi}_h$ and ${\theta}_h$. %To make the end-to-end framework differential, the quantization of the hyper latents is just the same as dealing the latents. 
For entropy coding the hyper latents, a factorized model is utilized to model the probability distribution. 

Following the work in~\cite{balle2018variational}, the authors in ~\cite{minnen2018joint} further proposed to model the probability distribution of the latents with a single Gaussian model, where the mean and scale parameters of the Gaussian distribution are estimated by a joint autoregressive and hyperprior. As shown in Fig. \ref{fig:fw3}, the estimation process is summarized as follows:
\begin{equation}
\begin{split}
p_{\hat{y}|\hat{z}} & \sim N(\mu,{\sigma}^2) \\
 with \  \mu,\sigma &= g_{ep}(\psi,{\phi}_i;{\theta}_{ep}) \\
\psi &=g_h(\hat{z};{\theta}_{hd}) \\
{\phi}_i &= g_{cm}(\hat{y}_{<i};{\theta}_{cm})
\end{split}
\end{equation}
where $g_{ep}$, $g_h$ and $g_{cm}$ represent the parameter estimation network, hyper synthesis transform and context estimation network, ${\theta}_{ep}$, ${\theta}_{hd}$ and ${\theta}_{cm}$ are the corresponding parameters, and $\psi$ and ${\phi}_i$ represent the information from the hyperprior and the local context. For the local context, 2D mask convolution is adopted to extract the spatial context, denoted as $C_S$.

For training the whole network, the training goal is to minimize the expected length of the bitstream as well as the expected distortion of the reconstructed image with respect to the original image, giving rise to a rate-distortion optimization problem:
\begin{equation}
\begin{split}
R + \lambda \cdot D &= E[-log_2(p_{\hat{y}|\hat{z}}({\hat{y}|\hat{z}}))] \\
& + E[-log_2(p_{\hat{z}}(\hat{z}))] + \lambda \cdot D(x,\hat{x})
\end{split}
\end{equation}
where the first and second terms represent the expected rate of the quantized latents and hyper latents,  and the third term represents the expected distortion of the reconstructed image with respect to the original image.

%-------------------------------------------------------------------------
\subsection{Motivation}

\begin{figure}[t]
\begin{center}
\subfigure[2D mask convolution]{
\includegraphics[width=0.90\linewidth]{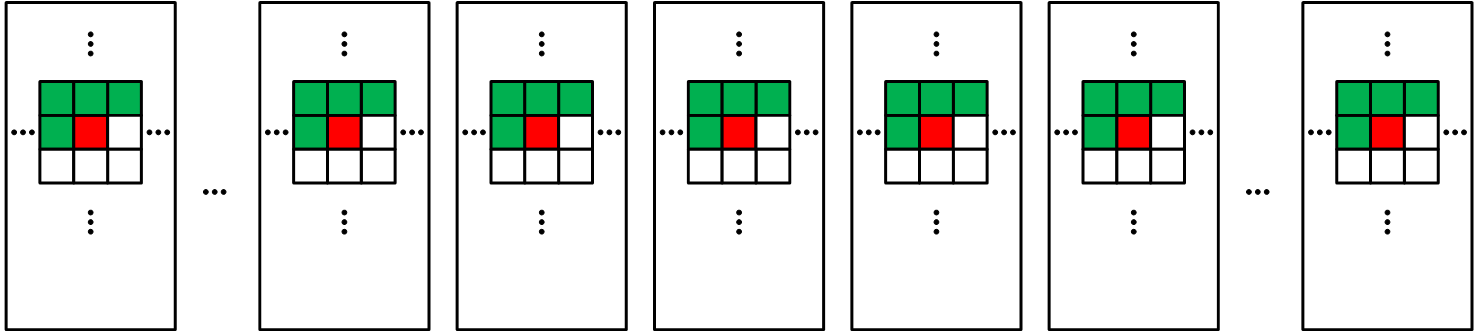}
\label{fig:2Dmask}
}
\subfigure[3D mask convolution]{
\includegraphics[width=0.90\linewidth]{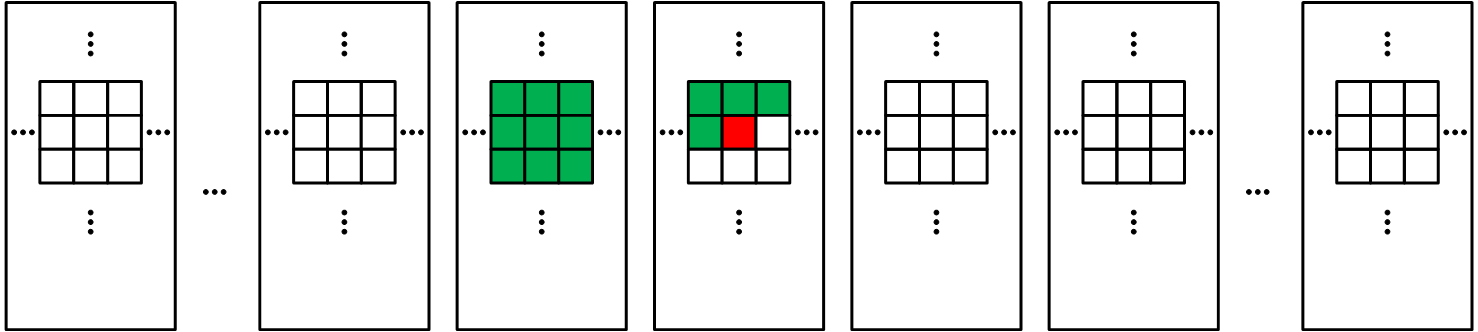}
\label{fig:3Dmask}
}
\subfigure[The proposed cross channel context model]{
\includegraphics[width=0.90\linewidth]{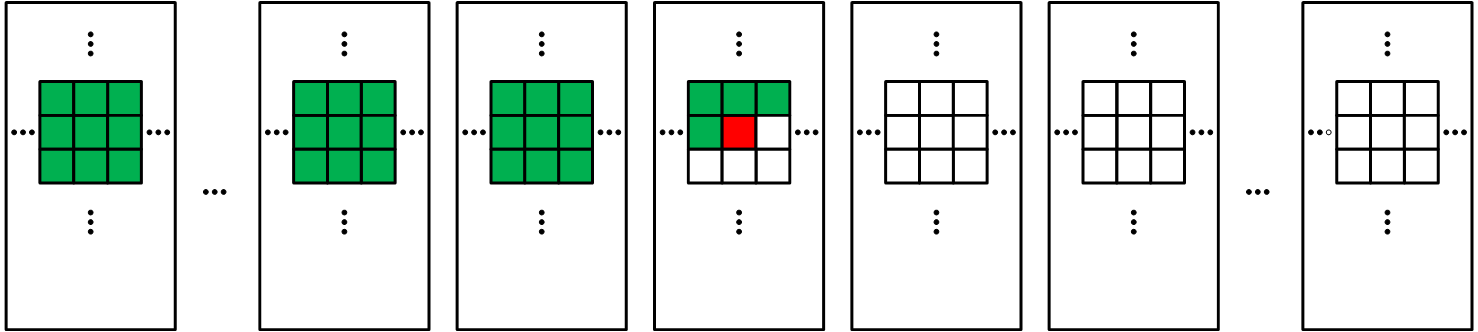}
\label{fig:CCCM}
}
\end{center}
   \caption{Different context models for latents in deep image compression, where the red blocks represent current latent elements, the green blocks represent the latent elements used as contexts for current latent elements.}
\label{fig:cm}
\end{figure}

\begin{figure}[t]
\begin{center}
\subfigure[Original image]{
\includegraphics[width=0.25\linewidth]{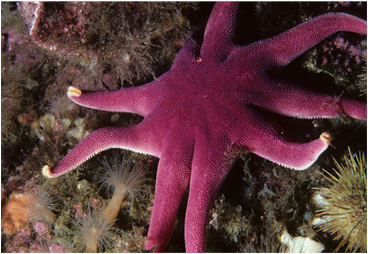}
\label{fig:0001}
}
\subfigure[Correlation comparison]{
\includegraphics[width=0.55\linewidth]{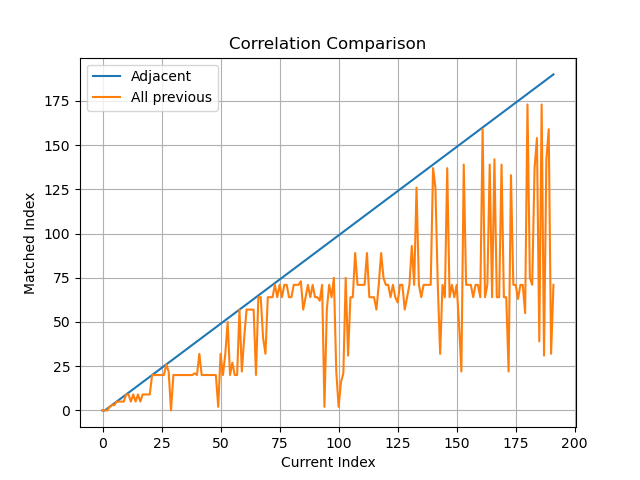}
\label{fig:CCIdx}
}
\subfigure[Example of correlation comparison]{
\includegraphics[width=0.85\linewidth]{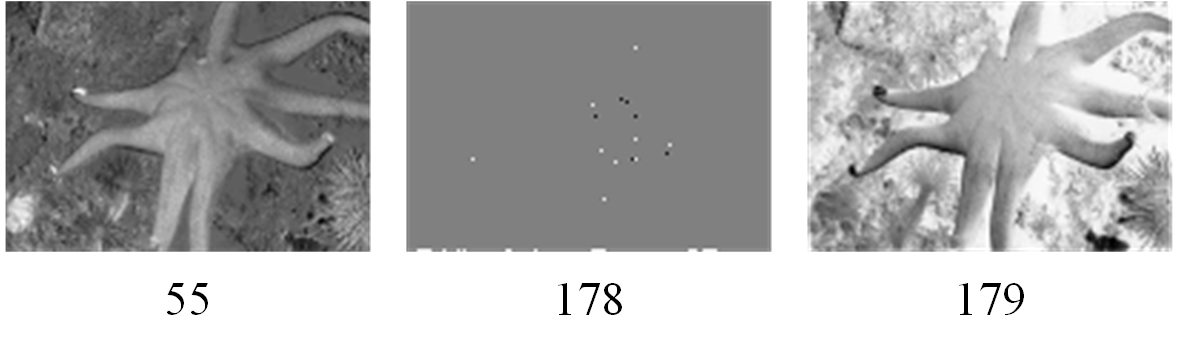}
\label{fig:group}
}
\end{center}
   \caption{Cross channel correlation analysis.}
\label{fig:exam}
\end{figure}

Currently, 2D mask convolution and 3D mask convolution are the main approaches to extract the local context from the quantized latents. In the following, we will analyze the drawbacks of 2D mask convolution and 3D mask convolution. Based on the analysis, we propose cross channel context model to address the shortcomings of the previous approaches.  

As shown in Fig. \ref{fig:2Dmask}, 2D mask convolution deals with latent elements from all channels at the same time. When using 2D mask convolution to capture the context information, the context latent elements are all on the top-left, top, top-right, and left of current latent elements. For one latent element, the co-located and nearby latent elements in previous channels with potentially strong correlations with the current latent element are not used as context for the current latent element. Generally speaking, 2D mask convolution mainly utilizes the causal spatial context information. %If the cross channel information can be further utilized, the coding performance may be improved.

As shown in Fig. \ref{fig:3Dmask}, when using 3D mask convolution to capture the context information, the context latent elements include those in previously coded adjacent channels and those causal adjacent latent elements in the current channel. However, there are two problems when using 3D mask convolution to capture both spatial and cross channel context information. 

First, different from spatial context where adjacent latent elements usually have strong correlations, the correlations between adjacent channels may or may not be strong as there is no constraints on the ranking of different latent channels when training the end-to-end image compression framework. To illustrate this, we use the pre-trained \emph{cheng2020-anchor} model by CompressAI~\cite{CompressAI} to compress the original image in Fig. \ref{fig:0001}, which comes from the DIV2K dataset~\cite{agustsson2017ntire}. After obtaining the quantized latents, we measure the correlations between two latent channels with mean absolute difference (MAD) of co-located elements. We compute the MAD of one channel with its all previous channels and select the channel with the minimum MAD as the matched channel. The orange line in Fig. \ref{fig:CCIdx} presents the matched channel results, where the blue line represents the adjacent previous channel. It can be observed that the matched channel varies widely, and most of the time is not the adjacent previous channel. As further evidence, we present an example in Fig. \ref{fig:group} with three successive images: the matched channel searched from all previous channels (channel index 55), the adjacent previous channel (channel index 178) and current channel (channel index 179). It can be clearly observed that, the matched channel has a significantly stronger correlation with the current channel than the adjacent previous channel, which verifies our idea.  

Second, the convolution kernel weights are shared between different channels when using 3D mask convolution. However, considering the correlations between different channels may be different, it may be better to use different convolution kernel weights to deal with different channels. 

To overcome the drawbacks of the 2D mask convolution and 3D mask convolution, as shown in Fig. \ref{fig:CCCM}, we propose the cross channel context model approach, where the latent elements in all previously coded channels together with the previously coded adjacent latent elements in the current channel are used as context for the current latent element. In addition, different convolution kernel weights are used to represent different correlations between different latent channels. 

It should be noted that our proposed method can be viewed as a more general 3D mask convolution. To generalize 3D mask convolution, the network needs to be carefully designed to support
asymmetric mask and deal several latents one time. In our paper, we choose a straightforward way to use 2D mask convolution to extract spatial context and 2D convolution to extract cross channel context, and we have
compared with the regular 3D mask convolution. 

\subsection{Cross Channel Context Model}

\begin{figure}[ht]
\begin{center}
\subfigure[Coding the first channel in the first group.]{
\includegraphics[width=0.92\linewidth]{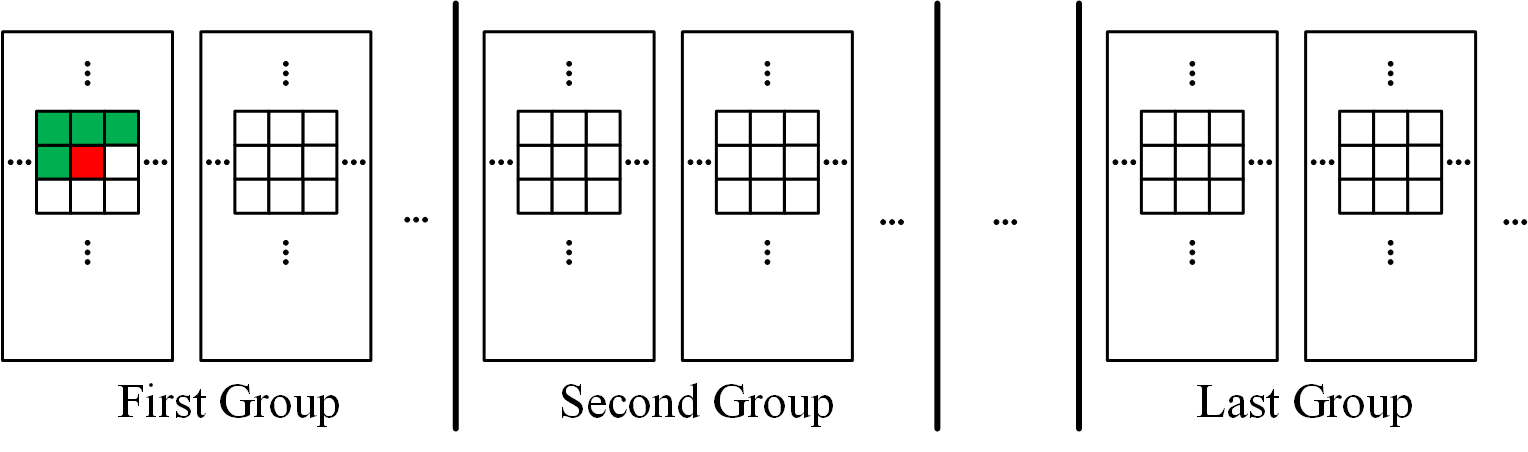}
\label{fig:step1}
}
\subfigure[Coding other channels in the first group.]{
\includegraphics[width=0.92\linewidth]{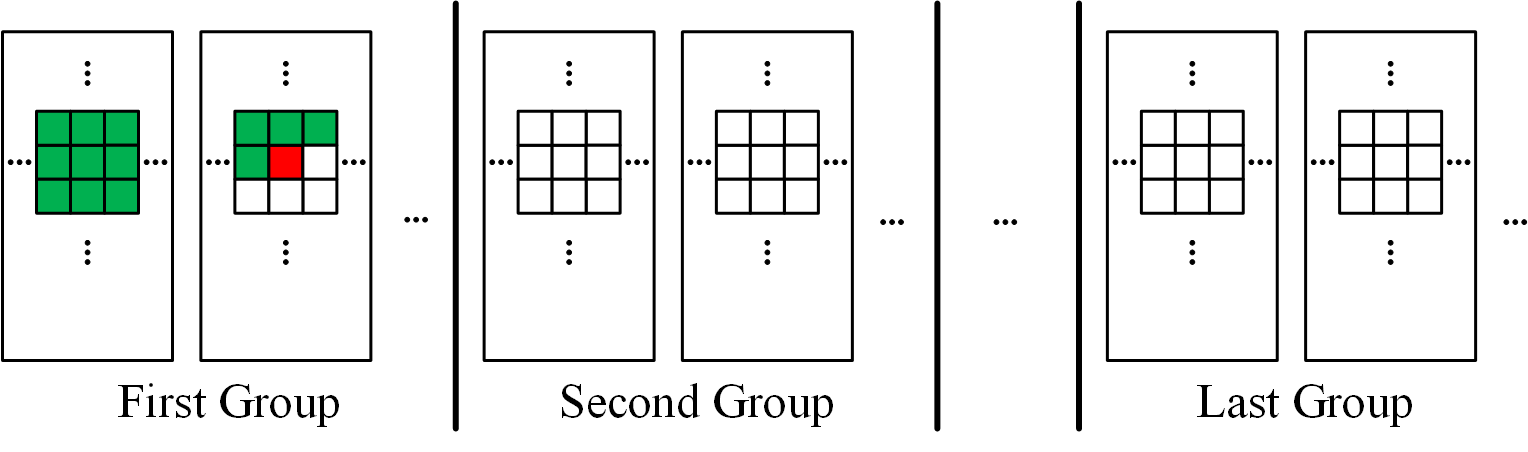}
\label{fig:step2}
}
\subfigure[Coding the second group.]{
\includegraphics[width=0.92\linewidth]{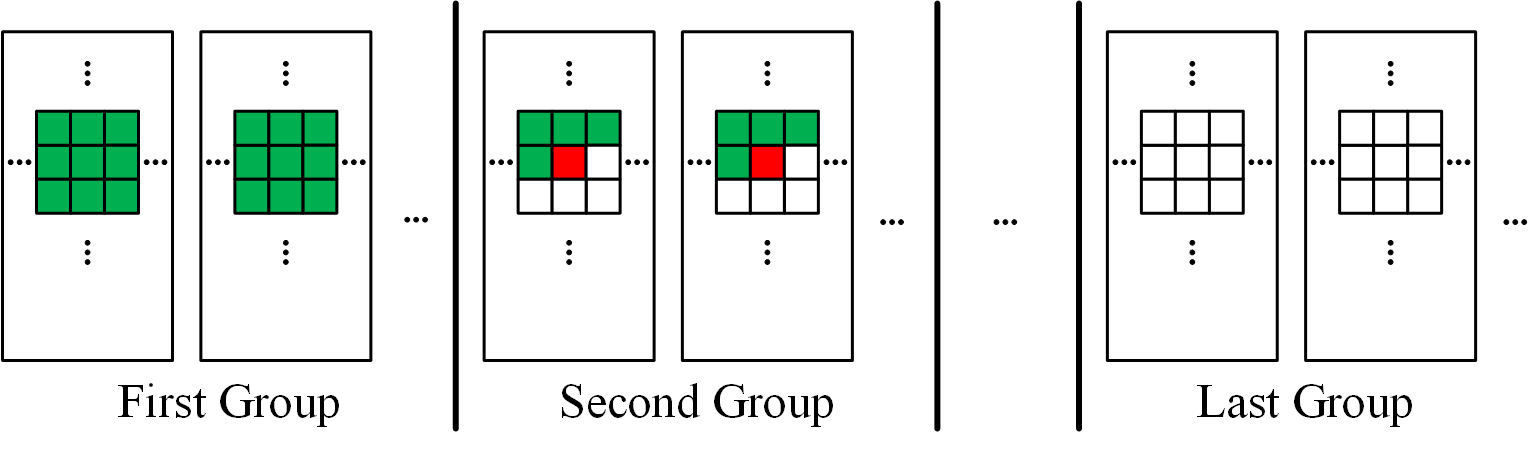}
\label{fig:step3}
}
\subfigure[Coding the last group.]{
\includegraphics[width=0.92\linewidth]{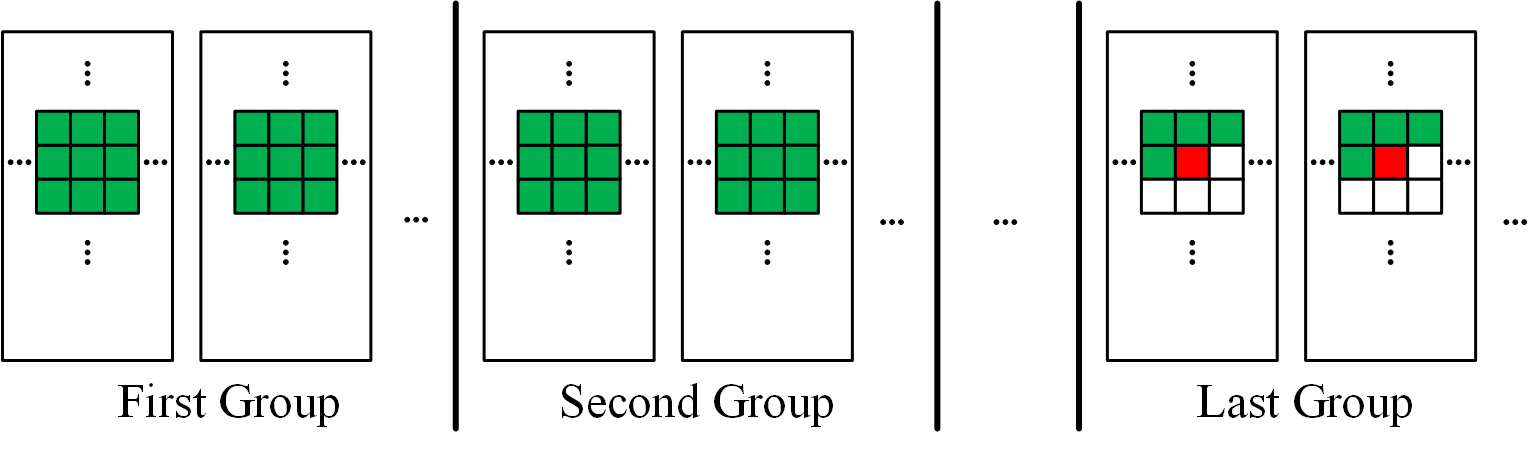}
\label{fig:step4}
}
\end{center}
   \caption{Basic steps of the proposed cross channel context model for Gaussian parameter estimation.}
\label{fig:steps}
\end{figure}

Based on the deep image compression framework with joint autoregressive and hyperprior model, the cross channel context, the spatial context, and the synthesis transformed hyper latents information are combined to estimate the Gaussian distribution parameters in the proposed model, as shown in Fig. \ref{fig:fw4}, where $C_{cc}$ represents cross channel context. Intuitively, to fully utilize the cross channel correlation, we should encode every channel one by one, and each channel should extract cross channel context information from all its previously encoded channels. However, considering the implementation complexity is too high, we divide the latents into several groups according to channel index. When encoding one group, all previously encoded groups can be utilized to provide the cross channel context information for current group. As there is no prior information about the distribution of the latents in channel dimension, we uniformly divide the latents into several groups.    

In our paper, we divide the latents into several groups according to channel index, such as channels with index 0, 1, ..., 15 are the first group, channels with index 16, 17, ..., 31 are the second group, and so on. In addition, we also tried to cluster or rank the channels during our research. For clustering, we attempted to divide the latents into several groups, assign one channel as cluster center for every group, with other latent channels referring the cluster center. For ranking, we attempted to use previously adjacent encoded channels to predict current channels. However, these attempts were not able to adaptively learn to cluster or rank. We will continue our research in this direction.

Fig. \ref{fig:steps} presents the basic steps of the proposed cross channel context model for Guassian parameter estimation. For the first channel in the first group ($\hat{y}_{11}$), as shown in Fig. \ref{fig:step1}, only spatial context (${\phi}_{si11}$) and the synthesis transformed hyper latents ($\psi_{11}$) information are combined to estimate the mean ($\mu_{11}$) and scale ($\sigma_{11}$) parameters of the first channel latents in the first group, as follows: 
%\begin{equation}
%\begin{split}
%p_{\hat{y}_{00}|\hat{z}} & \sim N(\mu_{00},{\sigma_{00}}^2) \\
 %with \  \mu_{00},\sigma_{00} &= g_{ep00}(\psi_{00},{\phi}_{si00};{\theta}_{ep00}) \\
%\psi_{00} &=g_{h00}(\hat{z};{\theta}_{hd00}) \\
%{\phi}_{si00} &= g_{cs00}(\hat{y}_{00<i};{\theta}_{cs00})
%\end{split}
%\end{equation}
\begin{equation}\label{equ5}
\begin{split}
p_{\hat{y}_{11}|\hat{z}} & \sim N(\mu_{11},{\sigma_{11}}^2) \\
 with \  \mu_{11},\sigma_{11} &= g_{ep11}(\psi_{11},{\phi}_{si11};{\theta}_{ep11}) \\
\psi_{11} &=g_{h11}(\hat{z};{\theta}_{hd11}) \\
{\phi}_{si11} &= g_{cs11}(\hat{y}_{11<i};{\theta}_{cs11})
\end{split}
\end{equation}
%, where $g_{h11}$, $g_{cs11}$ and $g_{ep11}$ represent the synthesis transform, spatial context estimation and entropy parameter estimation networks for the first channel in the first group, ${\theta}_{hd11}$, ${\theta}_{cs11}$ and ${\theta}_{ep11}$ are the corresponding parameters. 
For other channels in the first group ($\hat{y}_{12}$), as shown in Fig. \ref{fig:step2}, cross channel context (${\phi}_{cc12}$) is extracted from the first channel in the first group. Then, the extracted cross channel context, the spatial context (${\phi}_{si12}$) and the synthesis transformed hyper latents ($\psi_{12}$) information are combined to estimate the mean ($\mu_{12}$) and scale ($\sigma_{12}$) parameters of the other channels latents in the first group, as follows:  
%\begin{equation}
%\begin{split}
%p_{\hat{y}_{01}|\hat{z}} & \sim N(\mu_{01},{\sigma_{01}}^2) \\
 %with \  \mu_{01},\sigma_{01} &= g_{ep01}(\psi_{01},{\phi}_{si01},{\phi}_{cc01};{\theta}_{ep01}) \\
%\psi_{01} &=g_{h01}(\hat{z};{\theta}_{hd01}) \\
%{\phi}_{si01} &= g_{cs01}(\hat{y}_{00<i},\hat{y}_{01<i};{\theta}_{cs01}) \\
%{\phi}_{cc01} &= g_{cc01}(\hat{y}_{00};{\theta}_{cc01})
%\end{split}
%\end{equation}
\begin{equation}\label{equ6}
\begin{split}
p_{\hat{y}_{12}|\hat{z}} & \sim N(\mu_{12},{\sigma_{12}}^2) \\
 with \  \mu_{12},\sigma_{12} &= g_{ep12}(\psi_{12},{\phi}_{si12},{\phi}_{cc12};{\theta}_{ep12}) \\
\psi_{12} &=g_{h12}(\hat{z};{\theta}_{hd12}) \\
{\phi}_{si12} &= g_{cs12}(\hat{y}_{11<i},\hat{y}_{12<i};{\theta}_{cs12}) \\
{\phi}_{cc12} &= g_{cc12}(\hat{y}_{11};{\theta}_{cc12})
\end{split}
\end{equation}
%, where $g_{h12}$, $g_{cs12}$, $g_{cc12}$ and $g_{ep12}$ represent the synthesis transform, spatial context estimation, cross channel context estimation and entropy parameter estimation networks for other channels in the first group, ${\theta}_{hd12}$, ${\theta}_{cs12}$, ${\theta}_{cc12}$ and ${\theta}_{ep12}$ are the corresponding parameters. 
After coding the first group, for the second group ($\hat{y}_{2}$), as shown in Fig. \ref{fig:step3}, cross channel context (${\phi}_{cc2}$) is extracted from the first group ($\hat{y}_{1}$). Then, the extracted cross channel context (${\phi}_{cc2}$), the spatial context (${\phi}_{si2}$) and the synthesis transformed hyper latents ($\psi_2$) information are combined to estimate the mean ($\mu_2$) and scale ($\sigma_2$) parameters of the latents in the second group, as follows: 
%\begin{equation}
%\begin{split}
%p_{\hat{y}_1|\hat{z}} & \sim N(\mu_1,{\sigma_1}^2) \\
 %with \  \mu_1,\sigma_1 &= g_{ep1}(\psi_1,{\phi}_{si1},{\phi}_{cc1};{\theta}_{ep1}) \\
%\psi_1 &=g_{h1}(\hat{z};{\theta}_{hd1}) \\
%{\phi}_{si1} &= g_{cs1}(\hat{y}_{0<i},\hat{y}_{1<i};{\theta}_{cs1}) \\
%{\phi}_{cc1} &= g_{cc1}(\hat{y}_{0};{\theta}_{cc1})
%\end{split}
%\end{equation}
\begin{equation}\label{equ7}
\begin{split}
p_{\hat{y}_2|\hat{z}} & \sim N(\mu_2,{\sigma_2}^2) \\
 with \  \mu_2,\sigma_2 &= g_{ep2}(\psi_2,{\phi}_{si2},{\phi}_{cc2};{\theta}_{ep2}) \\
\psi_2 &=g_{h2}(\hat{z};{\theta}_{hd2}) \\
{\phi}_{si2} &= g_{cs2}(\hat{y}_{1<i},\hat{y}_{2<i};{\theta}_{cs2}) \\
{\phi}_{cc2} &= g_{cc2}(\hat{y}_{1};{\theta}_{cc2})
\end{split}
\end{equation}
%, where $g_{h2}$, $g_{cs2}$, $g_{cc2}$ and $g_{ep2}$ represent the synthesis transform, spatial context estimation, cross channel context estimation and entropy parameter estimation networks for the second group, ${\theta}_{hd2}$, ${\theta}_{cs2}$, ${\theta}_{cc2}$ and ${\theta}_{ep2}$ are the corresponding parameters.
For other groups, the corresponding mean and scale parameters are estimated in a similar manner. For example, for the last group ($\hat{y}_{n}$), as shown in Fig. \ref{fig:step4}, cross channel context (${\phi}_{ccn}$) is extracted from all the previously coded groups ($\hat{y}_{1},\hat{y}_{2},...,\hat{y}_{n-1}$). Then, the extracted cross channel context, the spatial context (${\phi}_{sin}$) and the synthesis transformed hyper latents ($\psi_{n}$) information are combined to estimate the mean ($\mu_{n}$) and scale ($\sigma_{n}$) parameters of the latents in the last group, as follows:
  
%\begin{equation}
%\begin{split}
%p_{\hat{y}_{n-1}|\hat{z}} & \sim N(\mu_{n-1},{\sigma_{n-1}}^2) \\
 %with \  \mu_{n-1},\sigma_{n-1} &= g_{ep(n-1)}(\psi_{n-1},{\phi}_{si(n-1)},{\phi}_{cc(n-1)};{\theta}_{ep(n-1)}) \\
%\psi_{n-1} &=g_{h(n-1)}(\hat{z};{\theta}_{hd(n-1)}) \\
%{\phi}_{si(n-1)} &= g_{cs(n-1)}(\hat{y}_{0<i},\hat{y}_{1<i},...,\hat{y}_{n-1<i};{\theta}_{cs(n-1)}) \\
%{\phi}_{cc(n-1)} &= g_{cc(n-1)}(\hat{y}_{0},\hat{y}_{1},...,\hat{y}_{n-2};{\theta}_{cc(n-1)})
%\end{split}
%\end{equation}
\begin{equation}\label{equ8}
\begin{split}
p_{\hat{y}_{n}|\hat{z}} & \sim N(\mu_{n},{\sigma_{n}}^2) \\
 with \  \mu_{n},\sigma_{n} &= g_{epn}(\psi_{n},{\phi}_{sin},{\phi}_{ccn};{\theta}_{epn}) \\
\psi_{n} &=g_{hn}(\hat{z};{\theta}_{hdn}) \\
{\phi}_{sin} &= g_{csn}(\hat{y}_{1<i},\hat{y}_{2<i},...,\hat{y}_{n<i};{\theta}_{csn}) \\
{\phi}_{ccn} &= g_{ccn}(\hat{y}_{1},\hat{y}_{2},...,\hat{y}_{n-1};{\theta}_{ccn})
\end{split}
\end{equation}
In equation \ref{equ5}, \ref{equ6}, \ref{equ7}, \ref{equ8},  $g_{hi}$, $g_{csi}$, $g_{ccj}$ and $g_{epi}$ ($i = 11, 12, 2, ...,n$, $j = 12, 2, ...,n$) represent the synthesis transform, spatial context estimation, cross channel context estimation and entropy parameter estimation networks, ${\theta}_{hdi}$, ${\theta}_{csi}$, ${\theta}_{ccj}$ and ${\theta}_{epi}$ are the corresponding parameters, which are different when coding different groups.

\subsection{Network Structure}

\begin{figure*}[t]
\begin{center}
\includegraphics[width=0.92\linewidth]{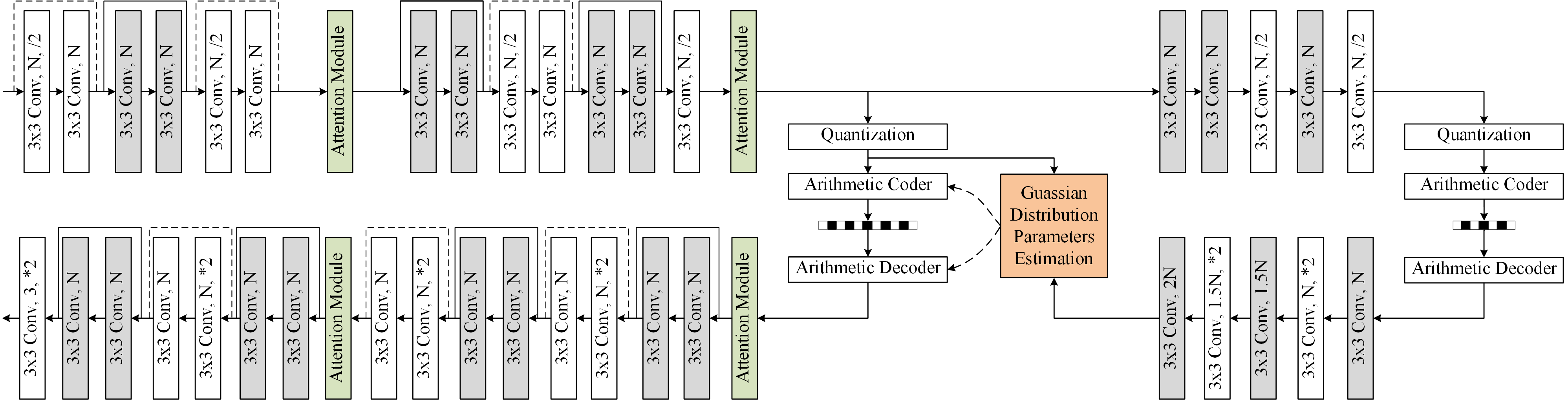}
\end{center}
\caption{Overall network architecture.}
\label{fig:network_all}
\end{figure*}

\begin{figure}[H]
\begin{center}
\includegraphics[width=0.65\linewidth]{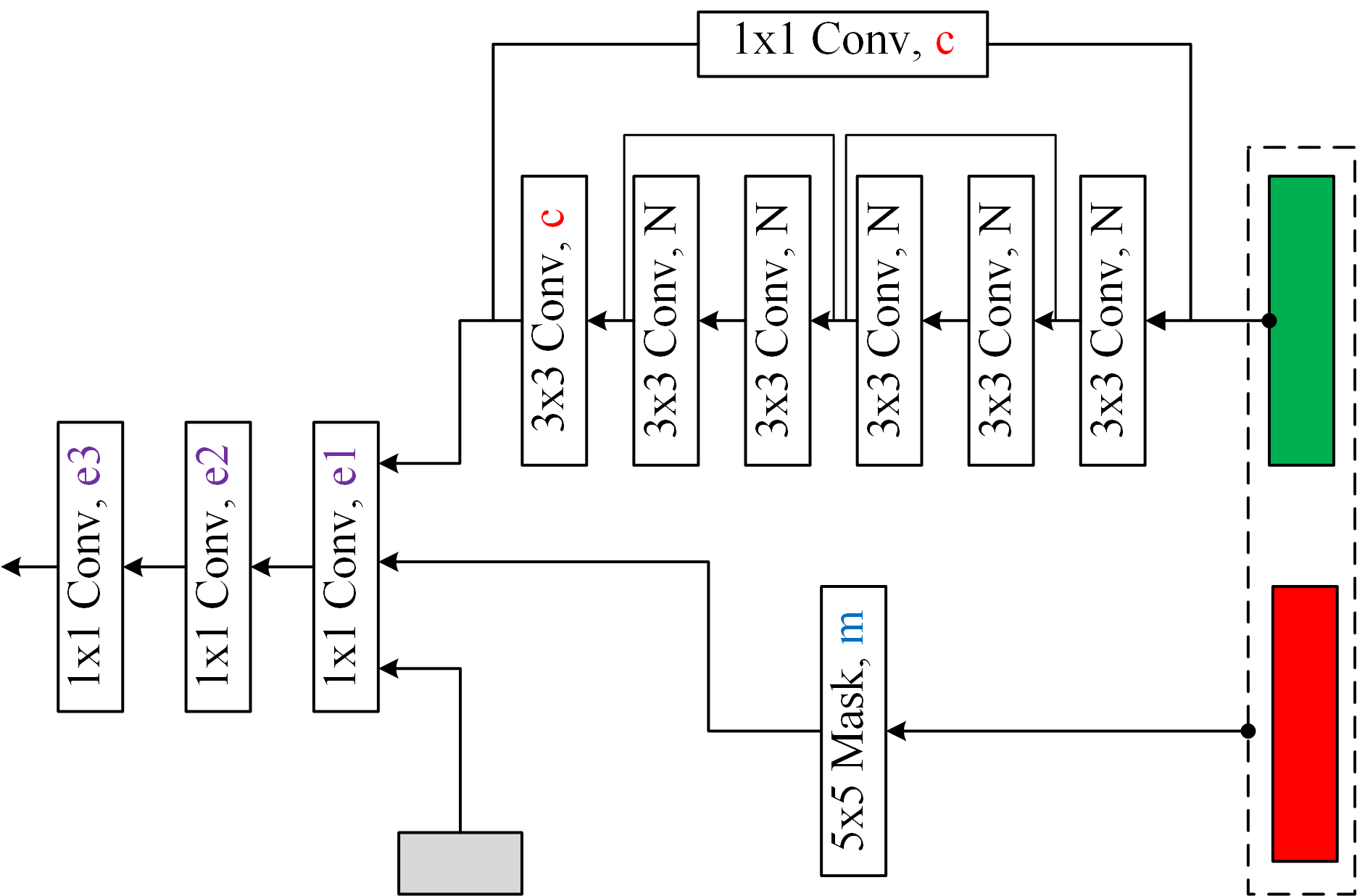}
\end{center}
\caption{Gaussian distribution parameters estimation network, where the green block represents previously coded groups, red block represents the group currently being coded, and gray block represents hyper synthesis transformed information.}
\label{fig:network_part}
\end{figure}

We integrate the proposed cross channel context model into the \emph{cheng2020-attn} model in CompressAI~\cite{CompressAI}, which is a simplified version of the model presented in ~\cite{cheng2020learned} with single Gaussian model. Fig. \ref{fig:network_all} presents the overall network structure. Our only change compared with the \emph{cheng2020-attn} model is the Gaussian distribution parameters estimation part, whose specific structure is presented in Fig. \ref{fig:network_part}.  In our implementation, we empirically uniformly divide the latents into 8 groups according to the channel index. When coding one group, the convolutional neural network with residual connection~\cite{he2016deep} is used to extract the cross channel context from previously coded groups (Green block in Fig. \ref{fig:network_part}), 2D mask convolution ($5\times5$ kernel size) is used to extract spatial context from both previously coded groups and the group currently being coded (Red block in Fig. \ref{fig:network_part}), $1\times1$ convolution kernel is used as the basic unit of the entropy parameter estimation network to fuse the cross channel context, the spatial context and hyper synthesis transformed information (Gray block in Fig. \ref{fig:network_part}) to estimate the Gaussian distribution parameters. 

\begin{table}
\begin{center}
\resizebox{\columnwidth}{!}
{
\begin{tabular}{l|c|c|c|c|c|c|c|c|c}
\hline
GIdx & 1$-$1 & 1$-$2 & 2 & 3 & 4 & 5 & 6 & 7 & 8\\
\hline%\hline
Ic & $-$ & 1 & N/8 & N/4 & 3N/8 & N/2 & 5N/8 & 3N/4 & 7N/8 \\
\hline
Is & 1 & N/8 & N/4 & 3N/8 & N/2 & 5N/8 & 3N/4 & 7N/8 & N \\
\hline
c  & $-$ & N/4$-$2 & N/4 & N/4 & N/4 & N/4 & N/4 & N/4 & N/4\\
\hline
m  & 2 & N/4$-$2 & N/4 & N/4 & N/4 & N/4 & N/4 & N/4 & N/4 \\
\hline
Id  & 2N & 2N & 2N & 2N & 2N & 2N & 2N & 2N & 2N \\
\hline
e1 & N$+$1 & 5N/4$-$2 & 5N/4 & 5N/4 & 5N/4 & 5N/4 & 5N/4 & 5N/4 & 5N/4 \\
\hline
e2 & N/2 & 5N/8$-$1 & 5N/8 & 5N/8 & 5N/8 & 5N/8 & 5N/8 & 5N/8 & 5N/8 \\
\hline
e3 & 2 & N/4$-$2 & N/4 & N/4 & N/4 & N/4 & N/4 & N/4 & N/4 \\
\hline
\end{tabular}
}
\end{center}
\caption{Specific parameter value.}
\label{tab:value}
\end{table}

Table \ref{tab:value} presents the specific parameter value used in our implementation. In this table, $Ic$ , $Is$ and $Id$ represent the channel number of the input to cross channel context estimation network, the input to spatial context estimation network, and hyper synthesis transformed information. $c$, $m$, $e1$, $e2$ and $e3$ as shown in Fig. \ref{fig:network_part} represent the channel number of the output of cross channel context estimation network ($c$), the output of spatial context estimation network ($m$) and the parameters of the entropy parameter estimation network ($e1$, $e2$ and $e3$). $N$ represents the channel number of the latents. $1-1$, $1-2$, $2$, ..., $8$ represent the first channel in the first group, other channels in the first group, the second group, ..., the eighth group, respectively.     

\subsection{Implementation Details}
\textbf{Training Details} The DIV2K dataset~\cite{agustsson2017ntire} and UCID dataset~\cite{schaefer2003ucid} are used in training. The networks are trained on $256\times256$ image patches randomly cropped from the images in the training datasets. The networks are trained using Adam~\cite{kingma2014adam} with batch size set to 16. The initial learning rate is set to $1\times10^{-4}$ for approximately $7\times10^5$ iterations, then the learning rate is reduced to $5\times10^{-5}$ for last approximately $3\times10^5$ iterations.

The networks are trained with two distortion metrics: mean squared error (MSE) and multi-scale structural similarity (MS-SSIM)~\cite{wang2003multiscale}. When optimized with MSE, we train 6 models with $\lambda$ value setting to \{0.018, 0.0035, 0.0067, 0.0130, 0.0250, 0.0483\}, the corresponding numbers of latent channels are \{128, 128, 128, 192, 192, 192\}. When optimized with MS-SSIM, the distortion term is defined as $D(x,\hat{x}) = 1 - $MS-SSIM$(x,\hat{x})$. We train 6 models with $\lambda$ value setting to \{4.58, 8.73, 16.64, 31.73, 60.50, 115.37\}, and the corresponding numbers of latent channels are \{128, 128, 192, 192, 192, 192\}.
  
\textbf{Testing Details} The Kodak dataset~\cite{franzen1999kodak} which contains 24 uncompressed $768\times512$ images and the CVPR workshop CLIC2020 professional dataset~\cite{cvprclic2020} which contains 250 2K images are used as the test sets. To evaluate the rate distortion performance, the rate is measured with bits per pixel (bpp), and the distortion is measured with PSNR or MS-SSIM depending on the training distortion metric. The rate-distortion (RD) curves are used to compare the coding performance of different methods. In addition, we also evaluate the specific coding performance values with BD rate reduction~\cite{pateux2007excel}, which is widely used during the development of traditional video coding standards. 

%-------------------------------------------------------------------------
\section{Experiments}

%-------------------------------------------------------------------------
\subsection{Rate-distortion Performance}

%\begin{figure}[t]
%\begin{center}
%\includegraphics[width=0.48\linewidth]{fig/Kodak_PSNR3.png}
%\end{center}
%\caption{Performance comparison.}
%\label{fig:Comparison3}
%\end{figure}

\begin{figure*}[t]
\begin{center}
\subfigure[MS-SSIM on Kodak]{
\includegraphics[width=0.292\linewidth]{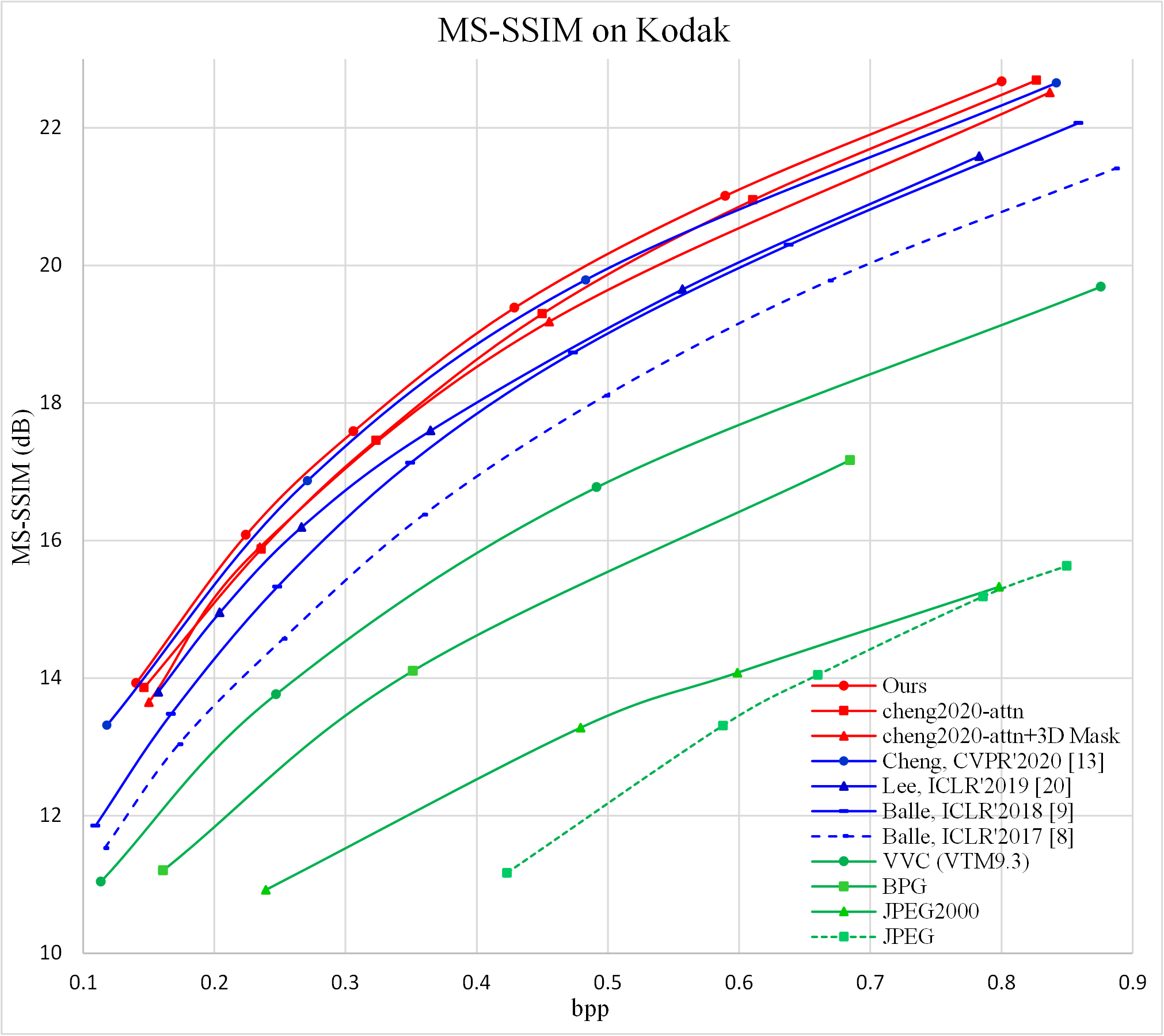}
\label{fig:rdkm}
}
\subfigure[PSNR on CLIC2020 Professional]{
\includegraphics[width=0.30\linewidth]{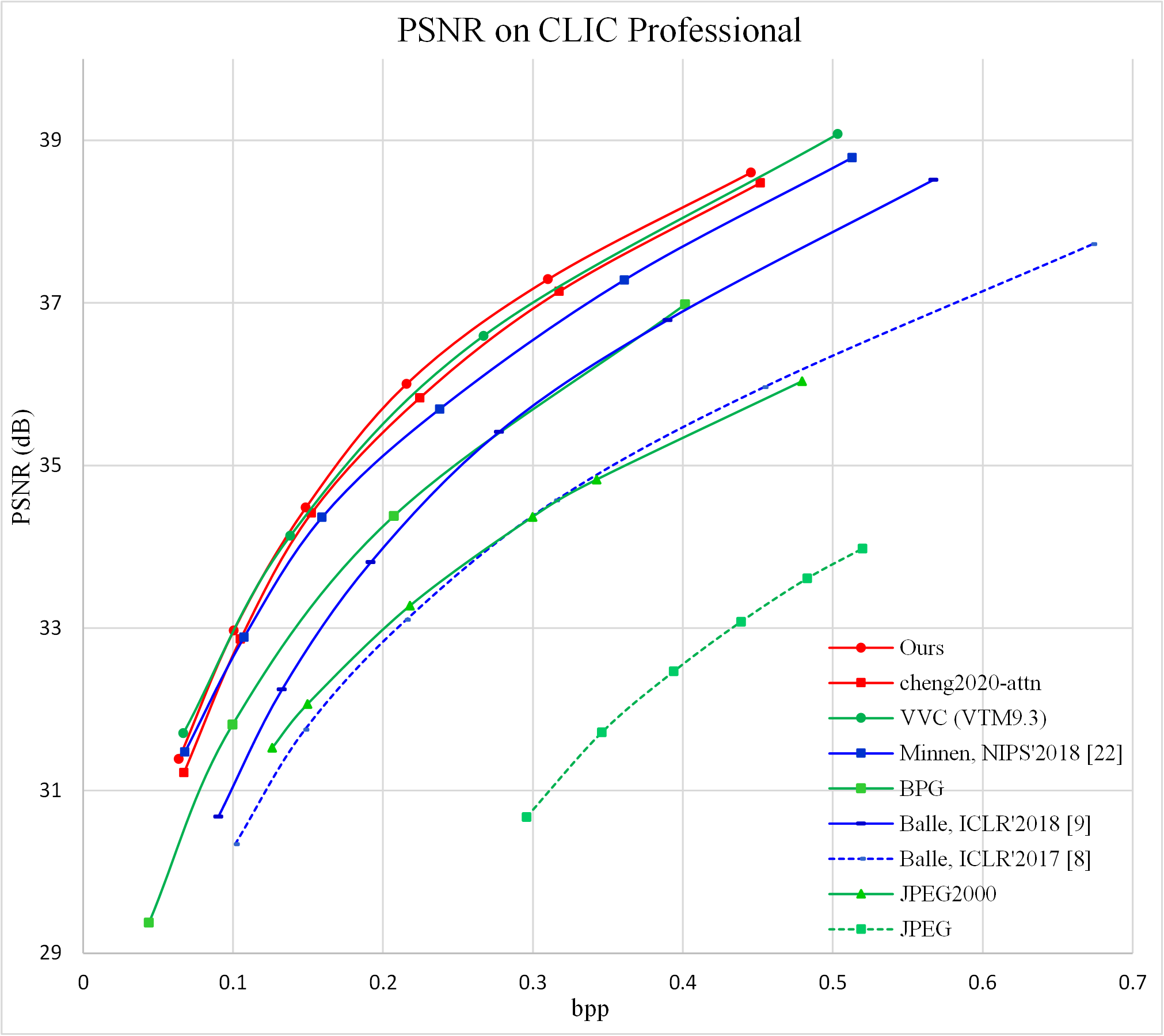}
\label{fig:rdcp}
}
\subfigure[MS-SSIM on CLIC2020 Professional]{
\includegraphics[width=0.30\linewidth]{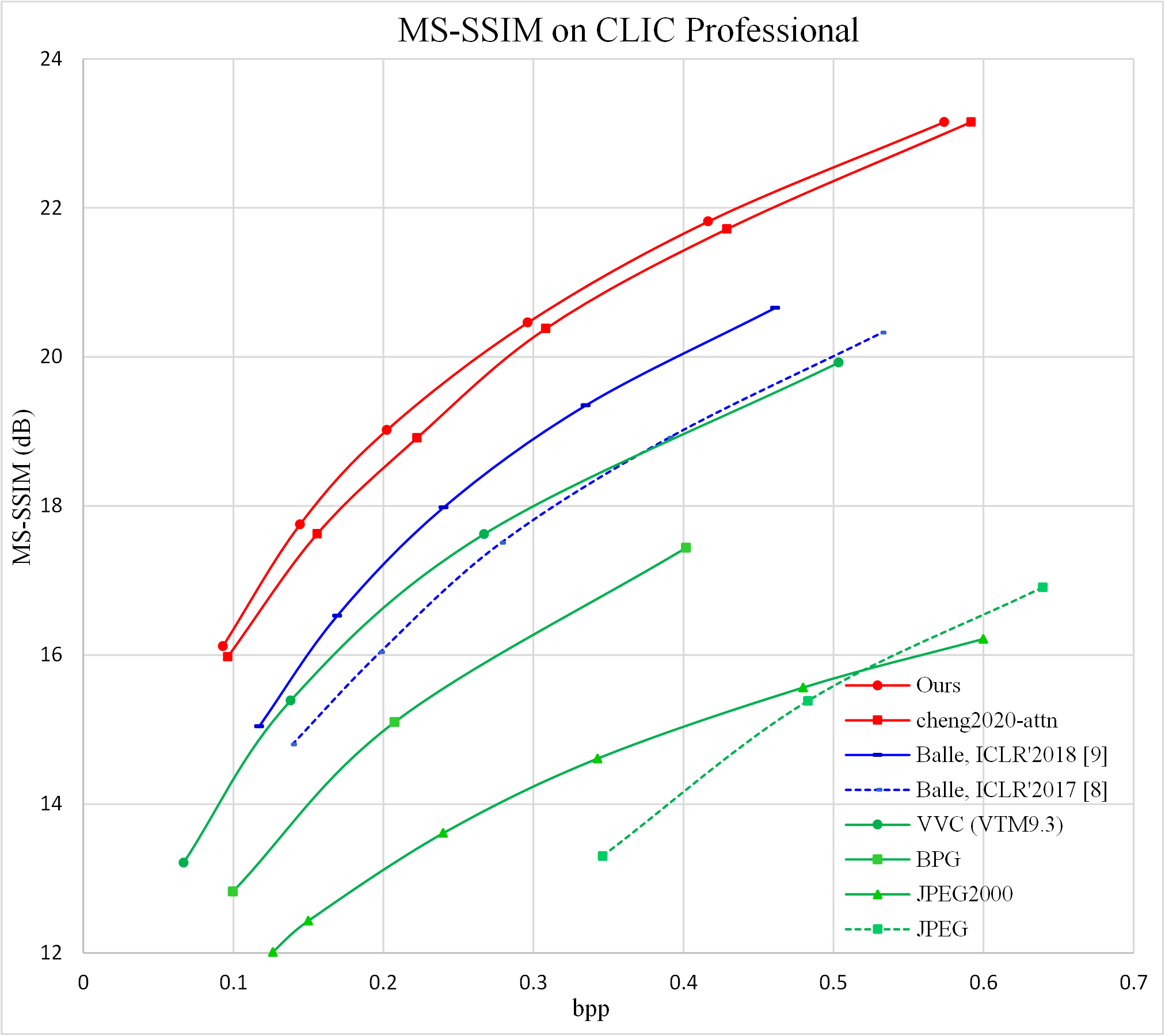}
\label{fig:rdcm}
}
\end{center}
   \caption{RD curves of different methods.}
\label{fig:rd}
\end{figure*}

We compare the proposed method with our baseline \emph{cheng2020-attn} model, \emph{cheng2020-attn} model integrated with 3D mask convolution, well-known compression standards, and recent deep image compression methods. The training details for \emph{cheng2020-attn} model and \emph{cheng2020-attn} model with 3D mask convolution are the same as our method, where the specific network structure for 3D mask convolution is the same as the implementation in~\cite{chen2019neural}. We compare to implementations of compression standards including VVC test model VTM9.3~\cite{VTM9.3}, BPG~\cite{bellard2015bpg}, JPEG2000~\cite{jpeg2000} and JPEG~\cite{JPEG}, and deep image compression methods including the work of Cheng \emph{et al.}~\cite{cheng2020learned}, Lee \emph{et al.}~\cite{lee2018context}, Minnen \emph{et al.}~\cite{minnen2018joint}, Ballé \emph{et al.}~\cite{balle2018variational} and Ballé \emph{et al.}~\cite{balle2016end}. For VTM9.3, we configure the software to code the images in YUV444 format. For Cheng \emph{et al.}~\cite{cheng2020learned}\footnote{\url{https://github.com/ZhengxueCheng/Learned-Image-Compression-with-GMM-and-Attention}}and Lee \emph{et al.}~\cite{lee2018context}\footnote{\url{https://github.com/JooyoungLeeETRI/CA_Entropy_Model}}, we use the results presented in their source code. For all the other methods, we use the results provided in CompressAI. 

Fig. \ref{fig:KodakPSNR} presents the RD curves on Kodak dataset in terms of PSNR distortion. Regarding PSNR, our proposed method achieves better coding performance than all the other methods, including VVC. %To our knowledge, our approach is the first work to achieve superior performance over VVC regarding PSNR.
When measured with BD-rate reduction, savings of 6.30\% over the baseline and 2.50\% over VVC have been achieved. It should be noted that \emph{cheng2020-attn} model integrated with 3D mask convolution doesn't achieve coding performance improvement compared with the baseline, the reason may be that 3D mask convolution can not effectively utilize the cross channel correlation, and the spatial correlation utilization is worse than 2D mask convolution.     

Fig. \ref{fig:rdkm} presents the RD curves on Kodak dataset with MS-SSIM distortion. For MS-SSIM, to show the difference more clearly, we convert it to $-10log_{10}(1-MS$-$SSIM)$. In terms of MS-SSIM, our proposed method achieves state-of-the-art coding performance. For CVPR CLIC2020 Professional dataset, Fig. \ref{fig:rdcp} and Fig. \ref{fig:rdcm} present the RD curves with PSNR and MS-SSIM distortion. In terms of PSNR, the proposed method achieves the best coding performance, and BD rate reductions of 6.31\% over the baseline and 2.20\% over VVC have been achieved. In terms of MS-SSIM, our proposed method also achieves state-of-the-art coding performance. Above results show that the proposed method is effective on low and high resolution images in terms of both PSNR and MS-SSIM distortions.
\subsection{Subjective Results}

%\begin{figure*}[t]
%\begin{center}
%\includegraphics[width=0.96\linewidth]{fig/kodim15.png}
%\end{center}
%\caption{Visualization of reconstructed images \emph{kodim15} from Kodak dataset.}
%\label{fig:kodim15}
%\end{figure*}

\begin{figure*}[t]
\begin{center}
\includegraphics[width=0.92\linewidth]{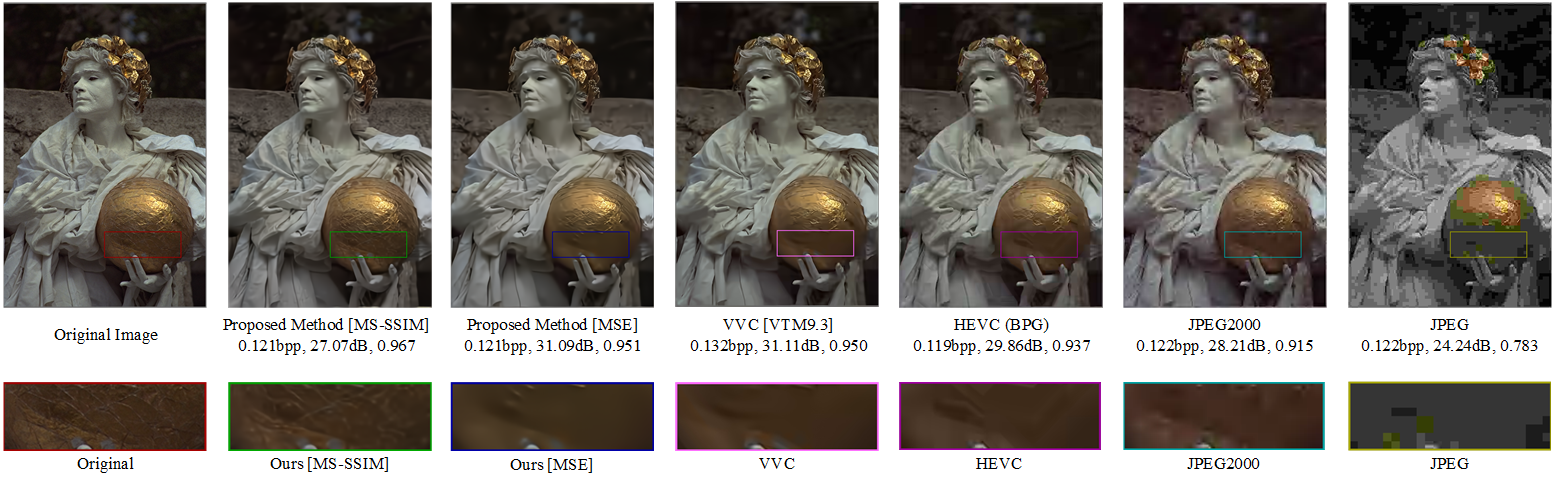}
\end{center}
\caption{Visualization of reconstructed images \emph{kodim17} from Kodak dataset.}
\label{fig:kodim17}
\end{figure*}

\begin{figure*}[t]
\begin{center}
\includegraphics[width=0.92\linewidth]{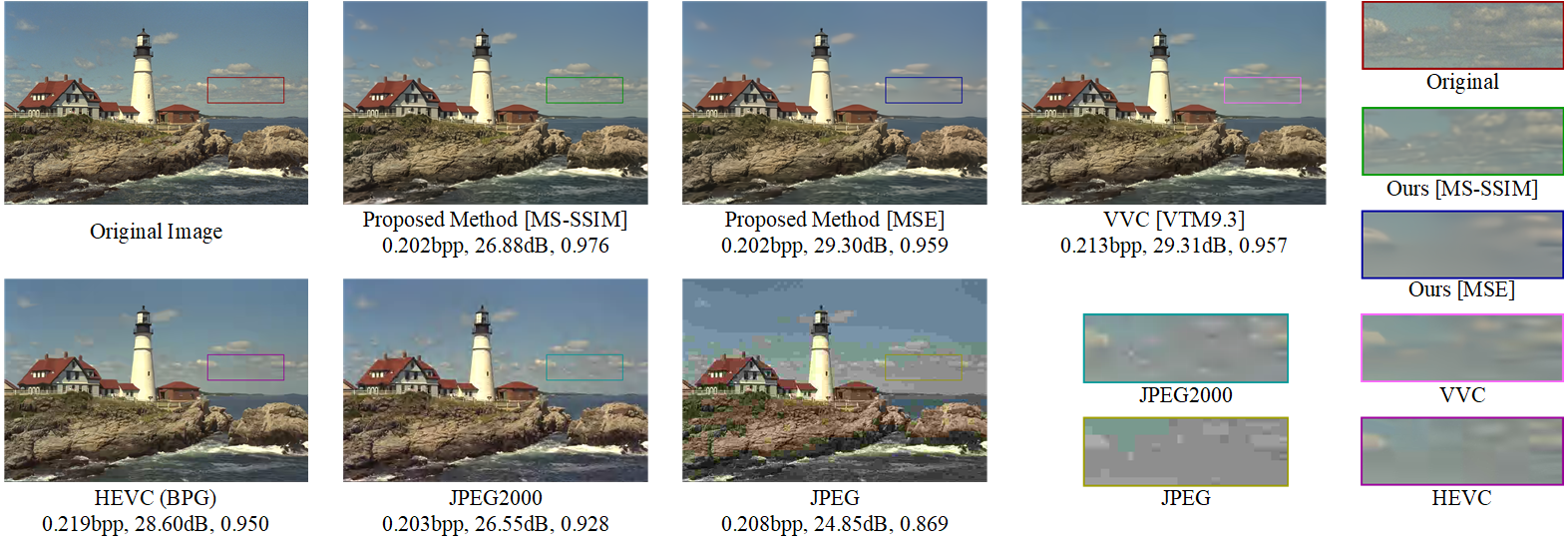}
\end{center}
\caption{Visualization of reconstructed images \emph{kodim21} from Kodak dataset.}
\label{fig:kodim21}
\end{figure*}

To demonstrate the proposed method can produce visually more pleasant results, we present some reconstructed images of different methods. 
%Fig. \ref{fig:kodim15} presents the reconstructed images of \emph{kodim15} with approximately 0.240bpp and compression ratio 100:1. It can be observed that the proposed method trained with MS-SSIM achieves the best visually pleasant reconstructed results compared with other methods, such as the mouth contains more texture information. The proposed method trained with MSE achieves a little better reconstructed results than VVC with less bits, and obviously outperforms BPG, JPEG2000, and JPEG. 
Fig. \ref{fig:kodim17} presents the reconstructed images of \emph{kodim17} with approximately 0.121 bpp and compression ratio of 198:1. It can be observed that the proposed method trained with MS-SSIM achieves the most visually pleasant reconstructed results compared with other methods. For example, in the enlarged portion containing the golden ball, more detailed texture is retained. The proposed method trained with MSE achieves comparable reconstructed results compared with VVC but with lower bit rates, and outperforms the reconstructed results of BPG, JPEG2000 and JPEG. Fig. \ref{fig:kodim21} presents the reconstructed images of \emph{kodim21} with approximately 0.202 bpp and compression ratio of 119:1. The proposed method trained with MS-SSIM retains the most texture, such as the cloud. In addition, the proposed method trained with MSE achieves comparable reconstructed results compared with VVC but with lower bit rates, and the reconstructed results of BPG, JPEG2000 and JPEG have obvious blurring or blocking artifacts.              

\subsection{Complexity Analysis}
As all the predicted channels can be compressed with one convolution layer, we take the number of serial operations as a main factor affecting the complexity, for latents with size $C \times H \times W$, the number for our paper is $9 \times H \times W$ (9: cross channel context, 8 groups with the first group divided into two parts; $H \times W$: spatial context), for 2D mask convolution is $H \times W$, for regular 3D mask convolution is $C \times H \times W $. For specific complexity value, the encoding time is about 2.45 times, and the decoding time is about 3.77 times compared with our baseline model. For specific parameter size, our model size is about 1.35 times compared with our baseline model. 

%------------------------------------------------------------------------
\section{Conclusion}
In this paper, we propose a cross channel context model for latents in deep image compression. The motivation comes from our observation that although there are strong correlations among different channels in latents, the strongest correlations may not always reside between adjacent channels. To effectively utilize the cross channel correlation, we divide the latents into several groups according to channel index, and the group currently being coded can utilize context information from all the previously coded groups to effectively capture the cross channel correlation. 

We integrate the proposed cross channel context model into the \emph{cheng2020-attn} model in CompressAI. Experimental results show that the proposed method can significantly improve the coding performance compared with the baseline. Furthermore, with the proposed method, the coding performance of deep image compression outperforms the latest video coding standard VVC when measured with PSNR distortion, which is a rare achievement. In addition, when trained with MS-SSIM, the proposed method also achieves more visually pleasant reconstructed results.      

{\small
\bibliographystyle{ieee_fullname}
\bibliography{cvpr}
}

\newpage
\begin{appendix}
%%%%%%%%% TITLE
%\title{A Cross Channel Context Model for Latents in Deep Image Compression \\ (Supplementary Material)}
%
%%\author{First Author\\
%%Institution1\\
%%Institution1 address\\
%%{\tt\small firstauthor@i1.org}
%%% For a paper whose authors are all at the same institution,
%%% omit the following lines up until the closing ``}''.
%%% Additional authors and addresses can be added with ``\and'',
%%% just like the second author.
%%% To save space, use either the email address or home page, not both
%%\and
%%Second Author\\
%%Institution2\\
%%First line of institution2 address\\
%%{\tt\small secondauthor@i2.org}
%%}
%
%\maketitle

%%%%%%%%% ABSTRACT
%\begin{abstract}
We provide more implementation details and experimental results in this supplementary material. Specifically, Section \ref{sec:sec1} presents more examples of correlation comparison. Section \ref{sec:sec2} presents the experimental results with different group division numbers. Section \ref{sec:sec3} presents the experimental results of \emph{cheng2020-attn} model integrated with 3D mask convolution in some CVPR CLIC2020 Professional images. Finally, Section \ref{sec:sec4} presents more reconstructed results of different methods for comparing the subjective results. 
%\end{abstract}

%%%%%%%%% BODY TEXT
\section{Correlation Comparison Examples}
\label{sec:sec1}

\begin{figure}[t]
\begin{center}
\includegraphics[width=0.96\linewidth]{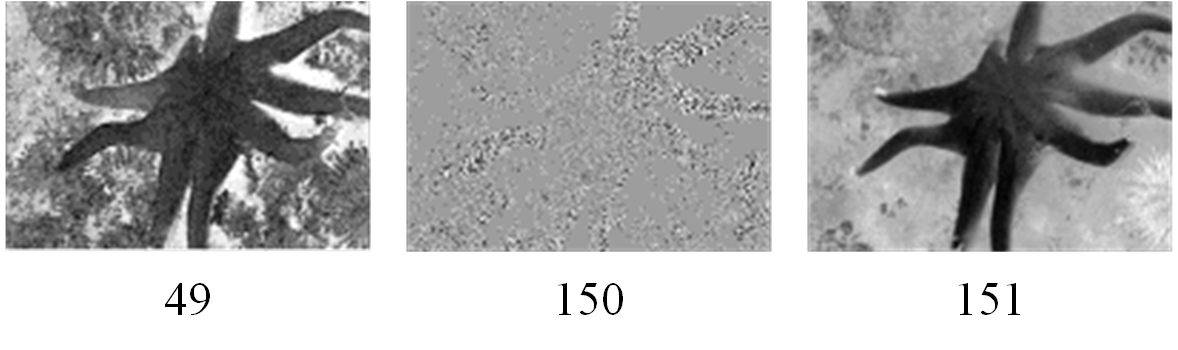}
\end{center}
\caption{Correlation comparison example 1.}
\label{fig:group2}
\end{figure}

\begin{figure}[t]
\begin{center}
\includegraphics[width=0.96\linewidth]{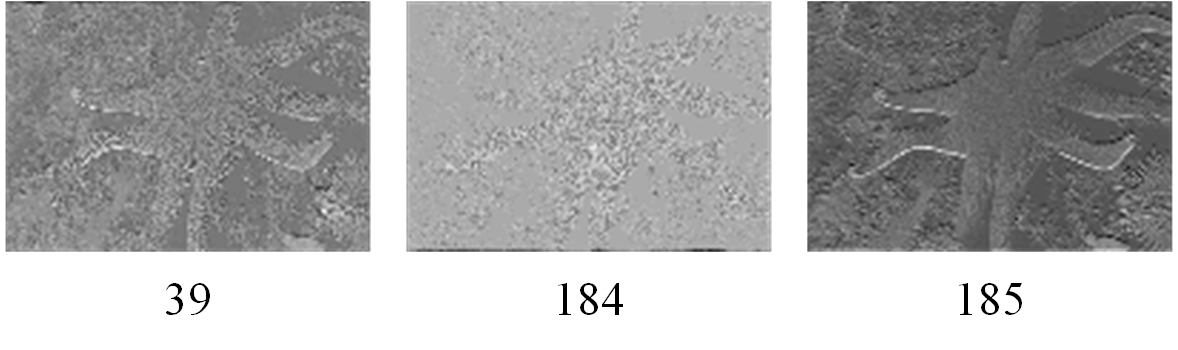}
\end{center}
\caption{Correlation comparison example 2.}
\label{fig:group3}
\end{figure}

We give two more correlation comparison examples in Fig. \ref{fig:group2} and Fig. \ref{fig:group3}. In these two figures, we successively present the matched channel searched from all previous channels, the adjacent previous channel and the current channel, with the latents presented above and the corresponding channel index presented below. It can be observed that there are strong correlations between different channels in the latents, notably the matched channel searched from all previous channels have a strong correlation with the current channel, but previously adjacent channel does not have a strong correlation. 

%-------------------------------------------------------------------------
\section{Experimental Results with Different Group Division Numbers}
\label{sec:sec2}

\begin{figure}[t]
\begin{center}
\includegraphics[width=0.96\linewidth]{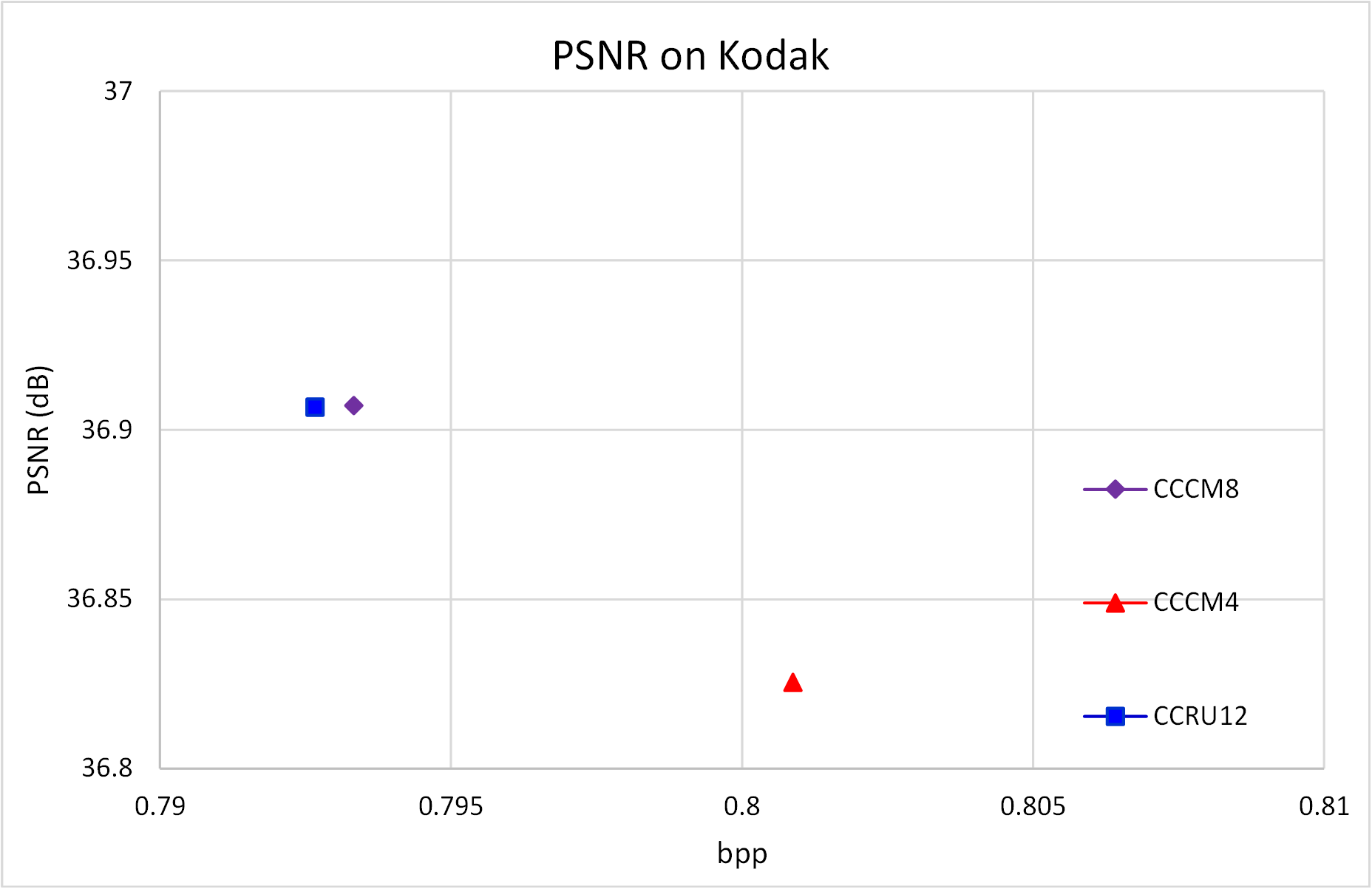}
\end{center}
\caption{Rate distortion performance on Kodak dataset with MSE distortion.}
\label{fig:RDpoint}
\end{figure}

In the paper, we empirically uniformly divide the latents into 8 groups according to the channel index. Here, we give the experimental results with 4 and 12 division groups. Fig. \ref{fig:RDpoint} presents the rate distortion points of different group division numbers on Kodak dataset with MSE distortion. It can be observed that when increasing the group division number from 4 to 8, the rate distortion performance has an obvious improvement. When further increasing the group division number from 8 to 12, the improvement becomes negligible. Considering more group division number means more complex implementation, we choose 8 as the default group division number.

\section{Experimental Results of \emph{cheng2020-attn} Model with 3D Mask Convolution}
\label{sec:sec3}

\begin{figure}[t]
\begin{center}
\includegraphics[width=0.96\linewidth]{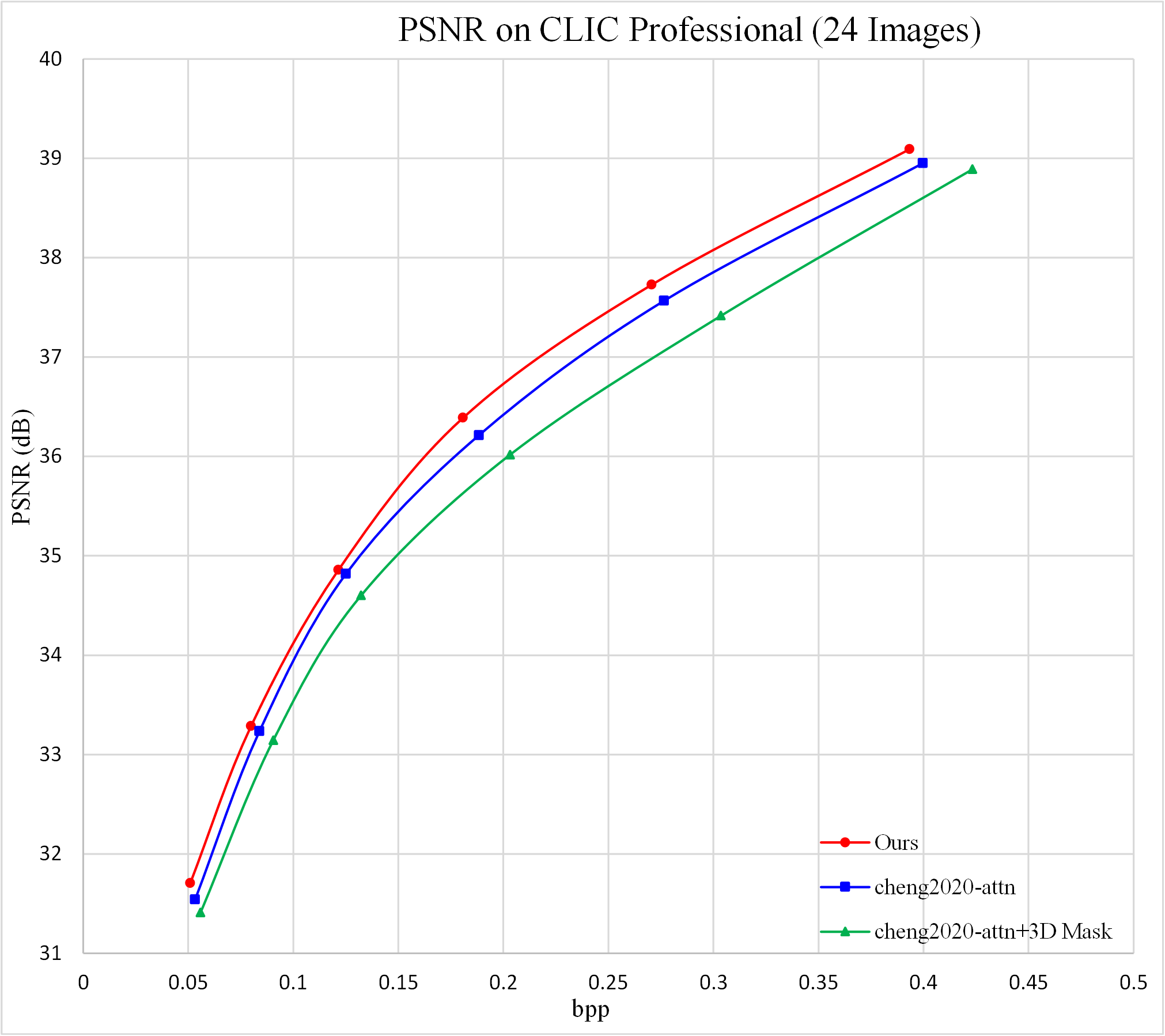}
\end{center}
\caption{RD Curves on 24 CLIC2020 professional images with PSNR distortion.}
\label{fig:3DMaskPSNR}
\end{figure}

\begin{figure}[t]
\begin{center}
\includegraphics[width=0.96\linewidth]{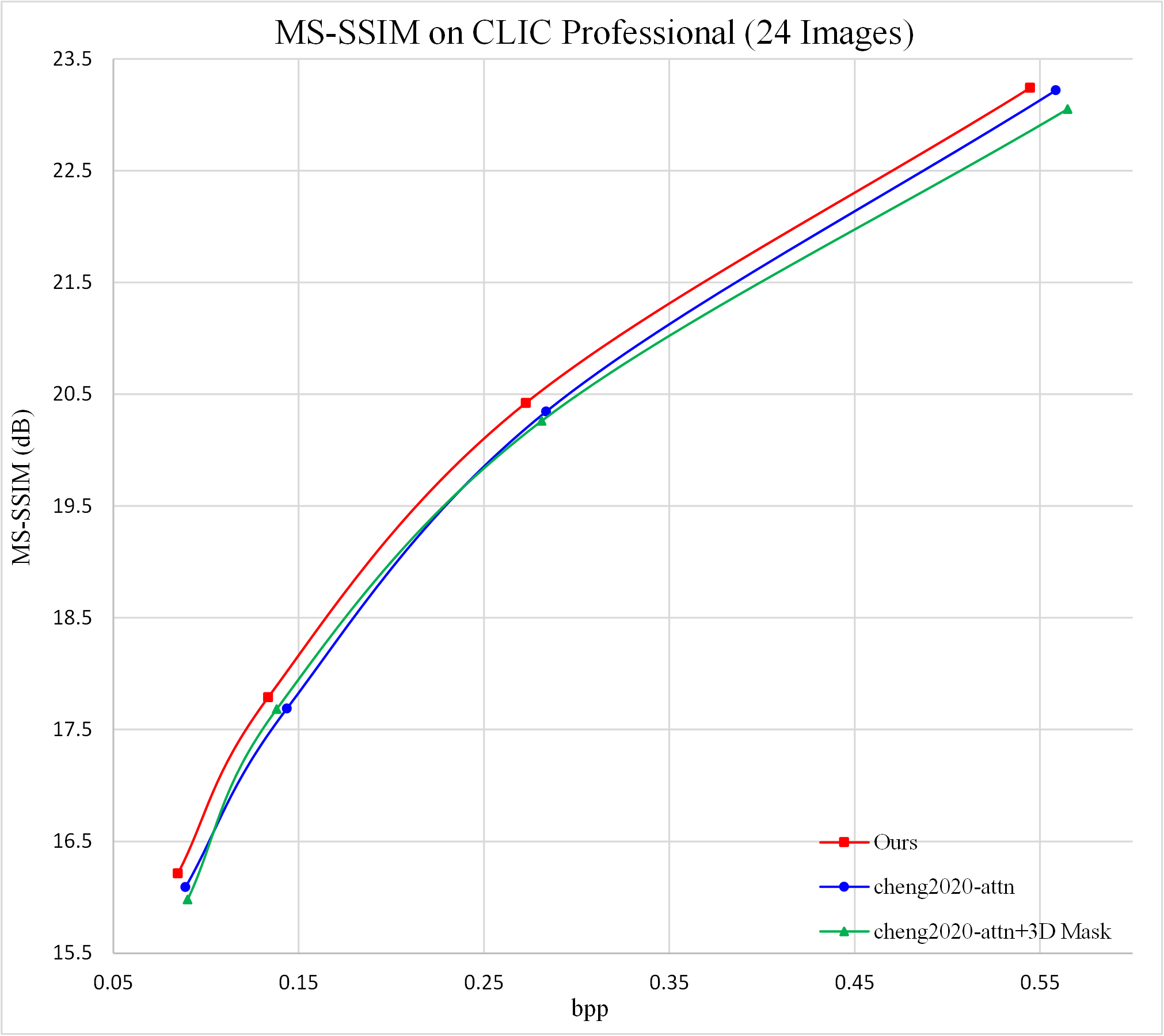}
\end{center}
\caption{RD Curves on 24 CLIC2020 professional images with MS-SSIM distortion.}
\label{fig:3DMaskMSSSIM}
\end{figure}

The performance of \emph{cheng2020-attn} model with 3D mask convolution is only compared with other methods in Kodak dataset in the paper. Here, we compare the coding performance of \emph{cheng2020-attn} model with 3D mask convolution, \emph{cheng2020-attn} model and our methods in part of the CVPR CLIC2020 professional dataset, where we choose the first 24 images in the dataset according to the image index. Fig. \ref{fig:3DMaskPSNR} and Fig. \ref{fig:3DMaskMSSSIM} present the RD curves of above three methods on 24 CLIC2020 professional images with PSNR and MS-SSIM distortion metric. It can be observed that \emph{cheng2020-attn} model integrated with 3D mask convolution doesn’t achieve coding performance improvement compared with the baseline. As a contract, our method obviously improves the coding performance compared with \emph{cheng2020-attn} model as it better utilizes the cross channel and spatial correlation in the latents.

%-------------------------------------------------------------------------
\section{Reconstructed Results of Different Methods}
\label{sec:sec4}

%-------------------------------------------------------------------------

%\begin{figure*}[t]
%\begin{center}
%\includegraphics[width=0.96\linewidth]{fig/kodim15.png}
%\end{center}
%\caption{Visualization of reconstructed images \emph{kodim15} from Kodak dataset.}
%\label{fig:kodim15}
%\end{figure*}

\begin{figure*}[t]
\begin{center}
\includegraphics[width=0.96\linewidth]{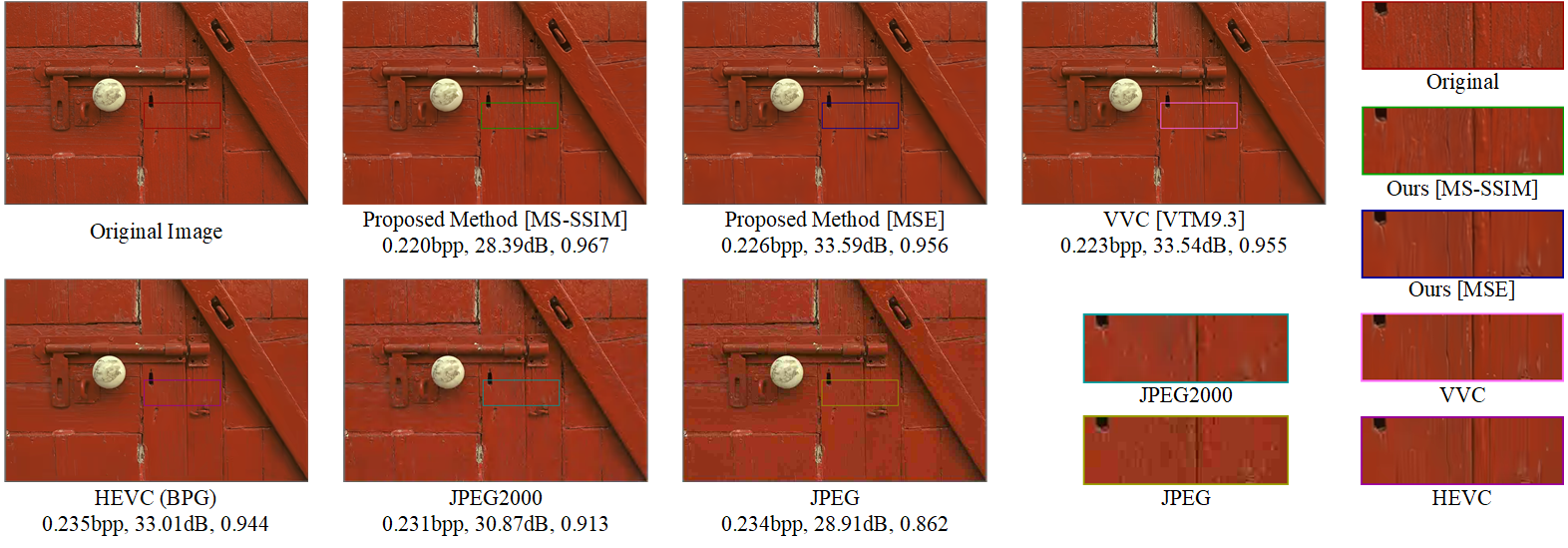}
\end{center}
\caption{Visualization of reconstructed images \emph{kodim02} from Kodak dataset.}
\label{fig:kodim02}
\end{figure*}

\begin{figure*}[t]
\begin{center}
\includegraphics[width=0.96\linewidth]{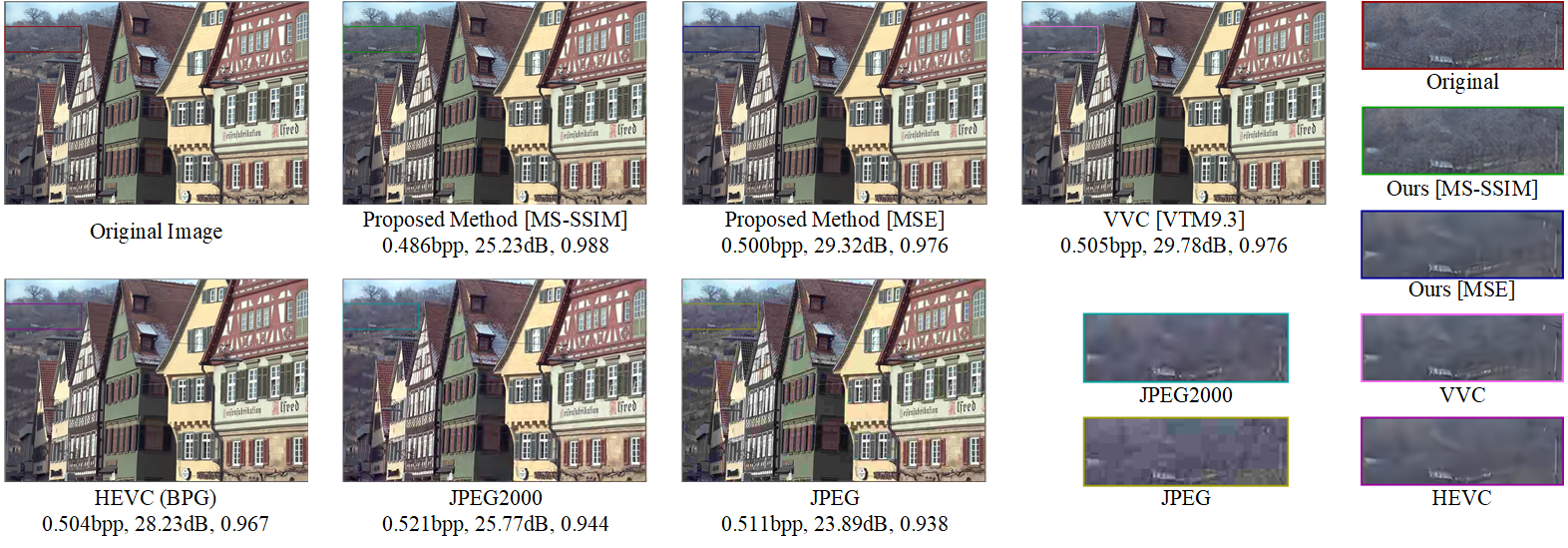}
\end{center}
\caption{Visualization of reconstructed images \emph{kodim08} from Kodak dataset.}
\label{fig:kodim08}
\end{figure*}

\begin{figure*}[t]
\begin{center}
\includegraphics[width=0.96\linewidth]{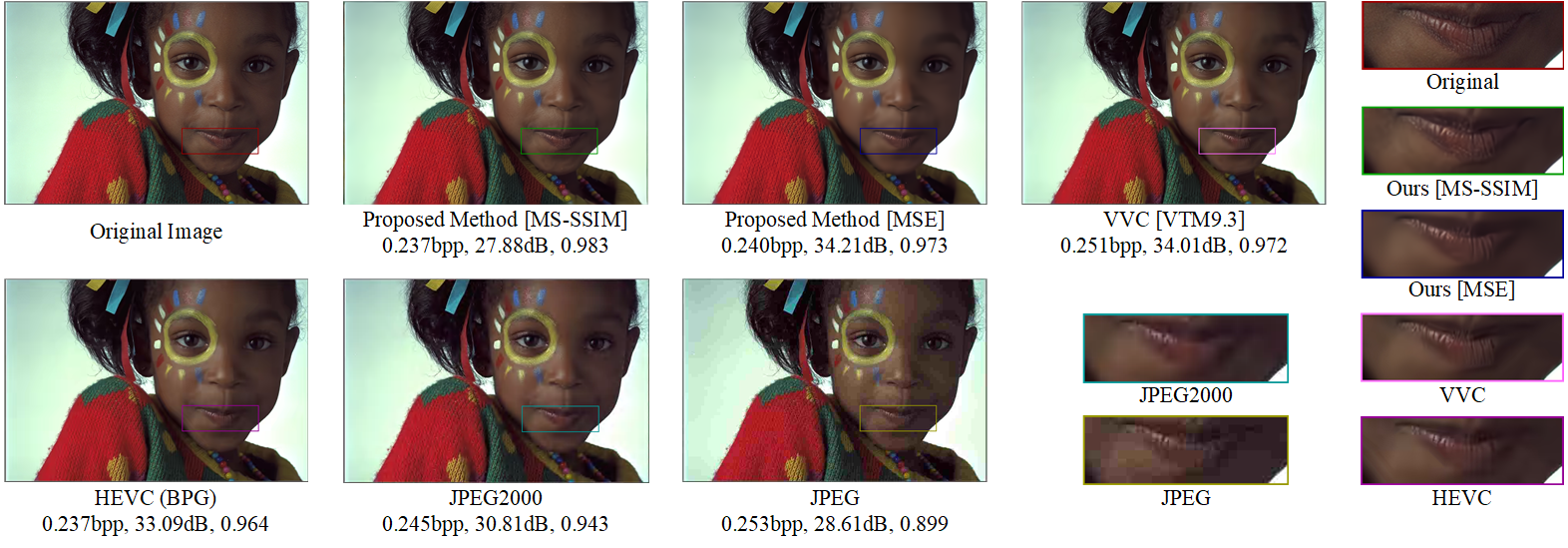}
\end{center}
\caption{Visualization of reconstructed images \emph{kodim15} from Kodak dataset.}
\label{fig:kodim15}
\end{figure*}

We give three more reconstructed images of different methods in Fig. \ref{fig:kodim02}, Fig. \ref{fig:kodim08}, and Fig. \ref{fig:kodim15}. Fig. \ref{fig:kodim02} presents the reconstructed images of \emph{kodim02} with approximately 0.220 bpp and compression ratio of 109:1. It can be observed that the proposed method trained with MS-SSIM achieves the most visually pleasant results. For example, the enlarged portion contains more texture information. The proposed method trained with MSE achieves reconstructed results comparable to VVC with approximately the same bit rates. For HEVC (BPG), JPEG2000, and JPEG, there are obvious blurring or blocking artifacts in the reconstructed images. Fig. \ref{fig:kodim08} presents the reconstructed images of \emph{kodim08} with approximately 0.486 bpp and compression ratio of 49:1. It can be observed that the proposed method trained with MS-SSIM retains the most texture information, such as the trees in the enlarged portion. The proposed method trained with MSE achieves reconstructed results comparable to VVC with slightly lower bit rates. For HEVC (BPG), JPEG2000, and JPEG, the reconstructed images have obvious blurring or blocking artifacts. Fig. \ref{fig:kodim15} presents the reconstructed images of \emph{kodim15} with approximately 0.240 bpp and compression ratio of 100:1. It can be observed that the proposed method trained with MS-SSIM achieves the most visually pleasant results. For example, the lip in the enlarged portion contains the most texture. The proposed method trained with MSE achieves slightly better reconstructed results than VVC with slightly lower bit rates, and outperforms the reconstructed results of HEVC (BPG), JPEG2000, and JPEG.

\end{appendix}

\end{document}